\documentclass[%
 reprint,
 amsmath,amssymb,
 aps,
]{revtex4-1}
\usepackage{blindtext}
\usepackage{comment}
\usepackage{subfigure}
\usepackage{graphicx}
\usepackage{dcolumn}
\usepackage{bm}
\usepackage{amssymb}
\usepackage{float}
\usepackage{graphicx}
\usepackage{mathtools}
\usepackage{subfigure}
\usepackage{type1cm}
\usepackage{eso-pic}
\usepackage{color}
\usepackage[scientific-notation=true]{siunitx}
\begin{document}
\title{
Flow regime transitions in dense non-Brownian suspensions:\\rheology, microstructural characterisation and constitutive modelling}
\author{Christopher Ness}
\author{Jin Sun}
\affiliation{University of Edinburgh, Edinburgh, EH9 3JL, United Kingdom}
\date{\today}
\begin{abstract}
Shear flow of dense, non-Brownian suspensions is simulated using the discrete element method, taking particle contact and hydrodynamic lubrication into account.
The resulting flow regimes are mapped in the parametric space of solid volume fraction, shear rate, fluid viscosity and particle stiffness.
Below a critical volume fraction $\phi_c$, the rheology is governed by the Stokes number, which distinguishes between viscous and inertial flow regimes.
Above $\phi_c$, a quasistatic regime exists for low and moderate shear rates.
At very high shear rates, the $\phi$ dependence is lost and soft particle rheology is explored.
The transitions between rheological regimes are associated with the evolving contribution of lubrication to the suspension stress.
Transitions in microscopic phenomena such as inter-particle force distribution, fabric and correlation length are found to correspond to those in the macroscopic flow. 
Motivated by the bulk rheology, a constitutive model is proposed combining a viscous pressure term with a dry granular model presented by Chialvo, Sun and Sundaresan [Phys. Rev. E. \textbf{85}, 021305 (2012)]. The model is shown to successfully capture the flow regime transitions.
\end{abstract}
\maketitle
\section{Introduction}

Dense suspensions of solid, non-Brownian particles in Newtonian fluid, such as slurries and pastes, are ubiquitous in nature and industry, and present a wealth of complex and surprising flow behavior~\cite{Stickel2005}.
Understanding the rheology of such suspensions is challenging, as the viscosity (shear stress divided by shear rate) is intimately linked to the solid volume fraction~\cite{Liu1998, Chialvo2012}, the shear rate, the preparation and shear history, and the particle properties.

Inspiration from granular mechanics~\cite{Forterre2008} has recently shed light on suspension rheology near the critical volume fraction, $\phi_c$.
Careful experimental work~\cite{Boyer2011a} demonstrates that dense suspensions can be constitutively characterised analogously to dry granular materials at low Reynolds numbers and below $\phi_c$ by adopting the popular $\mu(\text{I})$ rheology~\cite{Midi2004}. A suitable dimensionless control parameter, the viscous number $\text{I}_\text{V} = \eta_f \dot{\gamma} / \text{P}$ for wet systems with interstitial fluid viscosity $\eta_f$, confining pressure $\text{P}$ at a shear rate $\dot{\gamma}$  is defined, analogous to the much used inertial number $\text{I}_\text{I} = d\dot{\gamma} /\sqrt{\text{P}/\rho}$ \cite{Midi2004} for dry systems with particles of diameter $d$. Constitutive equations based on $\text{I}_\text{I}$ for the stress ratio $\mu = \sigma_{xy}/\text{P}$ and volume fraction have yielded striking matches with experimental results in some geometries for dry materials \cite{Jop2006}.
Computational work~\cite{Trulsson2012} has shown that the transition between dry and wet rheology below $\phi_c$ is continuous, and corresponds to a shift from particle contact to fluid dominated dissipation, hinting at a shift from Bagnoldian to Newtonian rheology at low shear rate and below $\phi_c$ as $\eta_f$ of a density matched, fully wetted, dense suspension is increased. In suspensions above $\phi_c$, experimental work has demonstrated that as the interstitial fluid viscosity is increased, the onset of a transition from arrested or quasistatic, rate-independent rheology to a flowing viscous regime is found to occur at decreasing shear rate~\cite{Huang2004}, though a particle-scale explanation of this behavior is missing.
While these works make considerable progress in bridging understanding between wet and dry systems, a complete picture of the rheology across flow regimes at both bulk and microscopic scales is lacking.

In the present work we shed light on the transitions between flow regimes with wet and dry characteristics, above and below $\phi_c$ for a broad range of shear rates and fluid viscosities. 
Below $\phi_c$, the rheology is governed by the Stokes number $\text{St} = \rho \dot{\gamma}d^2 /\eta_f$. The flow is viscous below $\text{St} = 1$, and quasi-Newtonian in the sense that the stress scales linearly with shear rate at constant volume fraction. For $\text{St} > 1$, particle inertia is important and the flow shear thickens, exhibiting Bagnoldian scaling of the shear stress. Such an inertial flow regime is highly reminiscent of dry granular flow, so by varying the Stokes number below $\phi_c$, we explore the transition from wet, fluid dominated rheology to ``dry'', particle contact dominated Bagnoldian rheology.
For volume fractions above $\phi_c$, the material has a yield stress and exhibits a quasistatic regime for low and moderate shear rates. Again, this behavior is reminiscent of arrested or jammed assemblies of slightly deformable dry grains.
For very high shear rates, particle overlaps become appreciable relative to the applied flow and the flow becomes less dependent on $\phi$. The rheology in this regime has been observed experimentally, for soft particle suspensions, to be intermediate (shear-thinning) \cite{Nordstrom2010a} or viscous \cite{Huang2004}, and we demonstrate that both behaviors can be captured computationally, dependent on the values of $\eta_f$ and the particle hardness $k_n$. 
The transitions between these flow regimes are shown to consistently correlate with the variation of the stress arising from hydrodynamic lubrication. Microscopic transitions in force and contact distributions and correlation length are also linked to the flow regime transitions. Finally, a constitutive model is developed to capture the transitions between regimes for all shear rates and volume fractions, serving as a unifying description.

The next section details the methods for solving particle dynamics, simulating simple shear flow and calculating bulk stresses. The bulk rheology and microstructural analysis are presented in Sections~\ref{sec:bulk} and \ref{sec:micro}, respectively. A constitutive model for the stresses in all flow regimes is proposed in Section~\ref{sec:const}, followed by a summary and concluding remarks in Section~\ref{sec:conc}.
\section{Numerical models and simulation details \label{sec:num}}

Discrete element method~\cite{Cundall1979} simulations are carried out using the particle simulation package LAMMPS~\cite{Plimpton1995}. The positions, velocities and forces of all particles are explicitly tracked over a period of time and are calculated in a step-wise, deterministic manner according to Newton's equations of motion. In accordance with recent works in dense suspension flow \cite{Trulsson2012,Seto2013}, we argue that in low-inertia, dense (volume fraction $\phi > 0.45$) suspension flows, the major fluid contribution to the stress can be effectively captured by resolving the normal and tangential pair-wise lubrication force~\cite{Kim1991,Ball1997} between neighbouring particles $i$ and $j$, with diameters $d_i$ and $d_j$ respectively, according to
\begin{subequations}
\begin{align}
&\mathbf{F}^{l,n}_{i} = \frac{3\pi\eta_f d^2_{ij}}{2} \frac{1}{h}(\textbf{v}_i - \textbf{v}_j) \cdot \textbf{n}_{ij} \textbf{n}_{ij}\text{,}\\
&
\begin{multlined}
\mathbf{F}^{l,t}_{i}  = \frac{4\pi\eta_f d_{ij}}{5} \left( 1 + \frac{d_{ij}}{d_i + d_j} + \left (\frac{d_i-d_j}{d_i+d_j} \right )^2 \right) \\ 
\ln \left({\frac{d_{i}}{2h}}\right) 
(\textbf{v}_i - \textbf{v}_j) \cdot (\mathbf{I}-\mathbf{n}_{ij}\mathbf{n}_{ij}) \text{,} 
\end{multlined}
\end{align}
\label{eq:lube}
\end{subequations}
for fluid viscosity $\eta_f$, weighted average particle diameter $d_{ij} = \frac{d_id_j}{d_i+d_j}$, surface-to-surface separation $h$, velocity vectors $\mathbf{v}_i$ and $\mathbf{v}_j$, center-to-center unit vector $\mathbf{n}_{ij}$ pointing from particle $j$ to $i$ and identity tensor $\mathbf{I}$. We assume that at sufficiently high volume fraction, the fluid in the narrow gaps between particles can be treated as laminar \cite{Lemaitre2009}. To limit computational expense and to mitigate the contact singularity of the lubrication force, $\mathbf{F}^{l,n}_{i}$ and $\mathbf{F}^{l,t}_{i}$ are calculated for $0.001 d_{ij} < h < 0.05 d_{ij}$. It has been verified that an outer cut-off of $0.1d_{ij}$ does not give significantly different results. We appeal to surface roughness to justify our choice of the inner cut-off. At smaller separations, we evaluate the forces assuming $h = 0.001d_{ij}$. 
For homogeneous, simple shear flow in a 3-dimensionally periodic domain, described later, we find that the dissipation arising due to lubrication dominates significantly over that arising from other fluid forces, such as Stokes drag and the Archimedes force described by \cite{Trulsson2012}, so we omit other such forces from this study.
Furthermore, we neglect fluid inertia from the model, assuming that (as demonstrated by \cite{Trulsson2012}) particle interactions will dominate the dissipation for inertial flows.

In addition to the lubrication force, particles $i$ and $j$ interact at contact ($h<0$) through a repulsive force, with normal and tangential components given by a linear spring-dashpot model for stiffness constants $k_n$ and $k_t$, center-to-center displacement $\delta_{ij}$, elastic shear displacement \textbf{u}$^t_{ij}$
\begin{subequations}
\begin{align}
\mathbf{F}^{c,n}_{ij} &=k_n \delta_{ij} \textbf{n}_{ij}  \text{,} \\
\mathbf{F}^{c,t}_{ij} &=-k_t \textbf{u}^t_{ij} \text{.}
\end{align}
\label{eq:cont}
\end{subequations}

A Coulomb friction coefficient $\mu_p$ is defined such that the tangential force on each particle is limited to $|\mathbf{F}^{c,t}{_{ij}}| \le \mu_p |\mathbf{F}^{c,n}{_{ij}}|$. We note that the damping arising from the lubrication force term is sufficient to achieve a steady state without employing a thermostat, so further damping in the mechanical contact model is omitted. The particle friction coefficient is fixed at $\mu_p=1$, which affects the critical volume fraction as discussed later. Variation of other particle parameters does not change the results presented in this paper. The solid and fluid phases are density matched, so no gravitational force is applied to the particles.

\begin{figure}
  \centering
      \subfigure[]{
  \includegraphics[height=37mm]{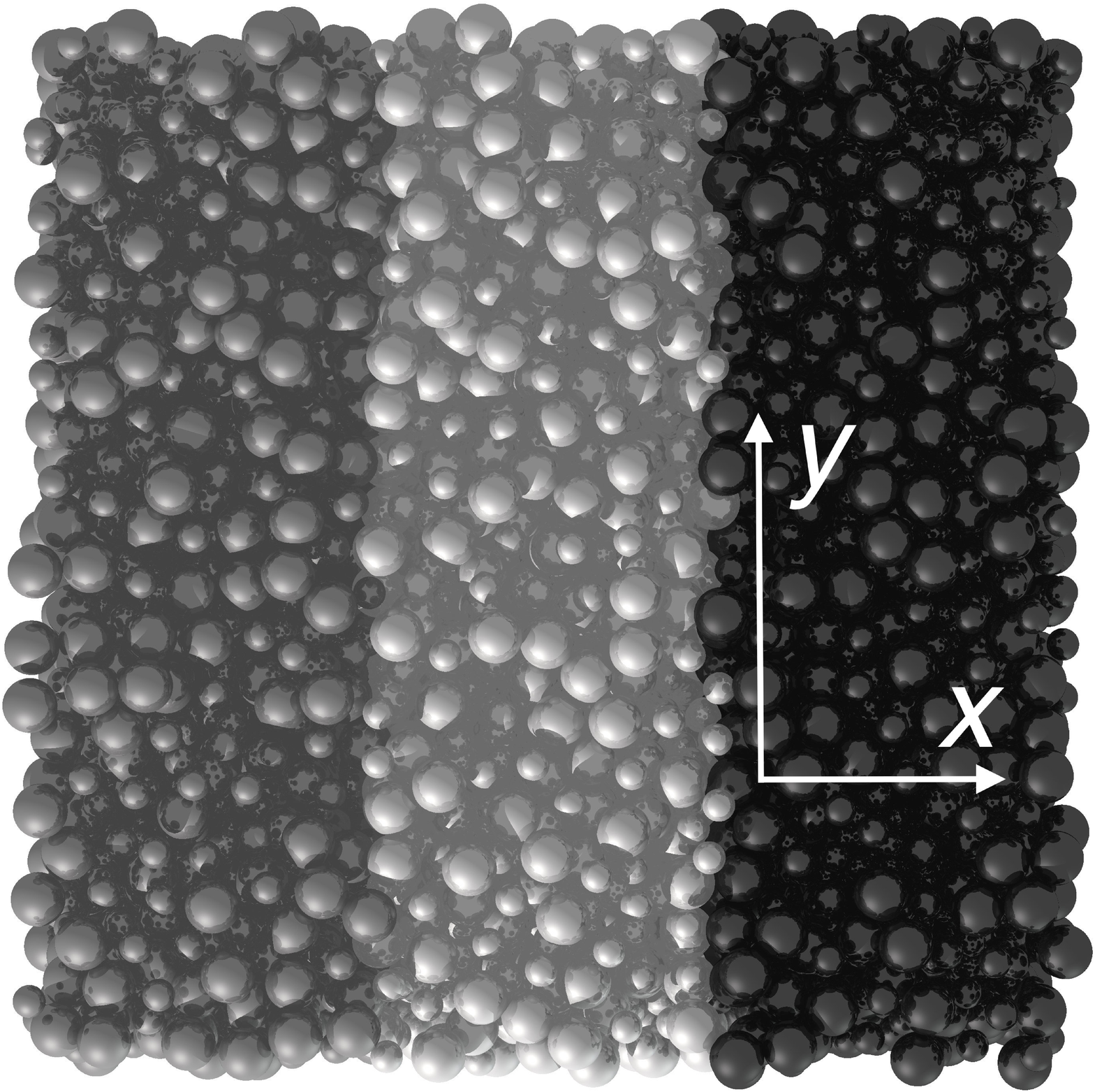}}
      \subfigure[]{
  \includegraphics[height=37mm]{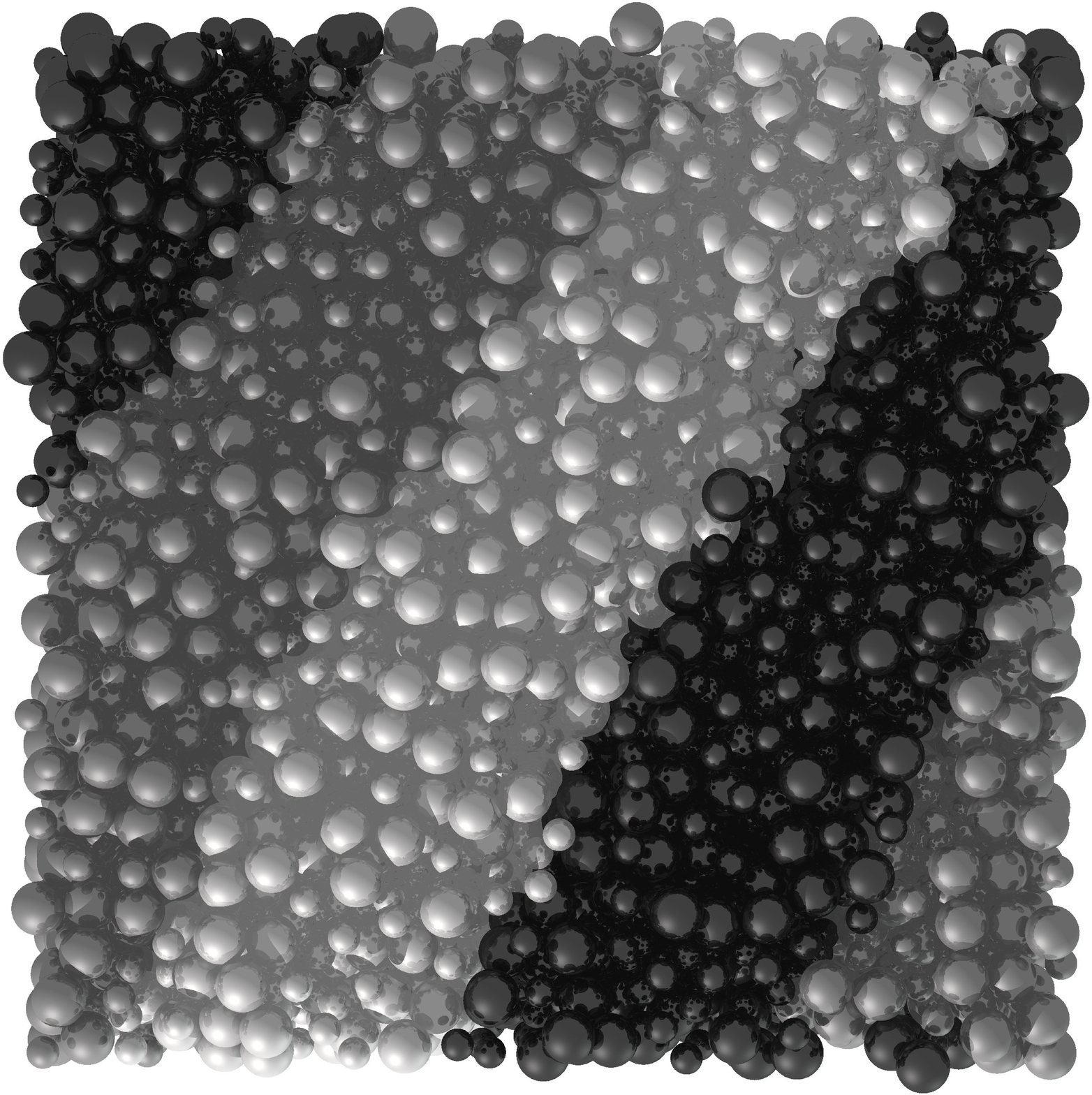}}
     \caption{
Snapshot of a particle assembly under steady shear at some time $t=0$ (a) and a later time $t=t_1$ (b), with a viewpoint normal to the $x$-$y$-plane where $x$ is the flow direction and $y$ is the velocity gradient direction. The colors are used to illustrate the deformation being applied to the assembly.
 }
 \label{fig:particles}
\end{figure}

\begin{figure*}
  \centering
      \subfigure[]{
  \includegraphics[height=65mm]{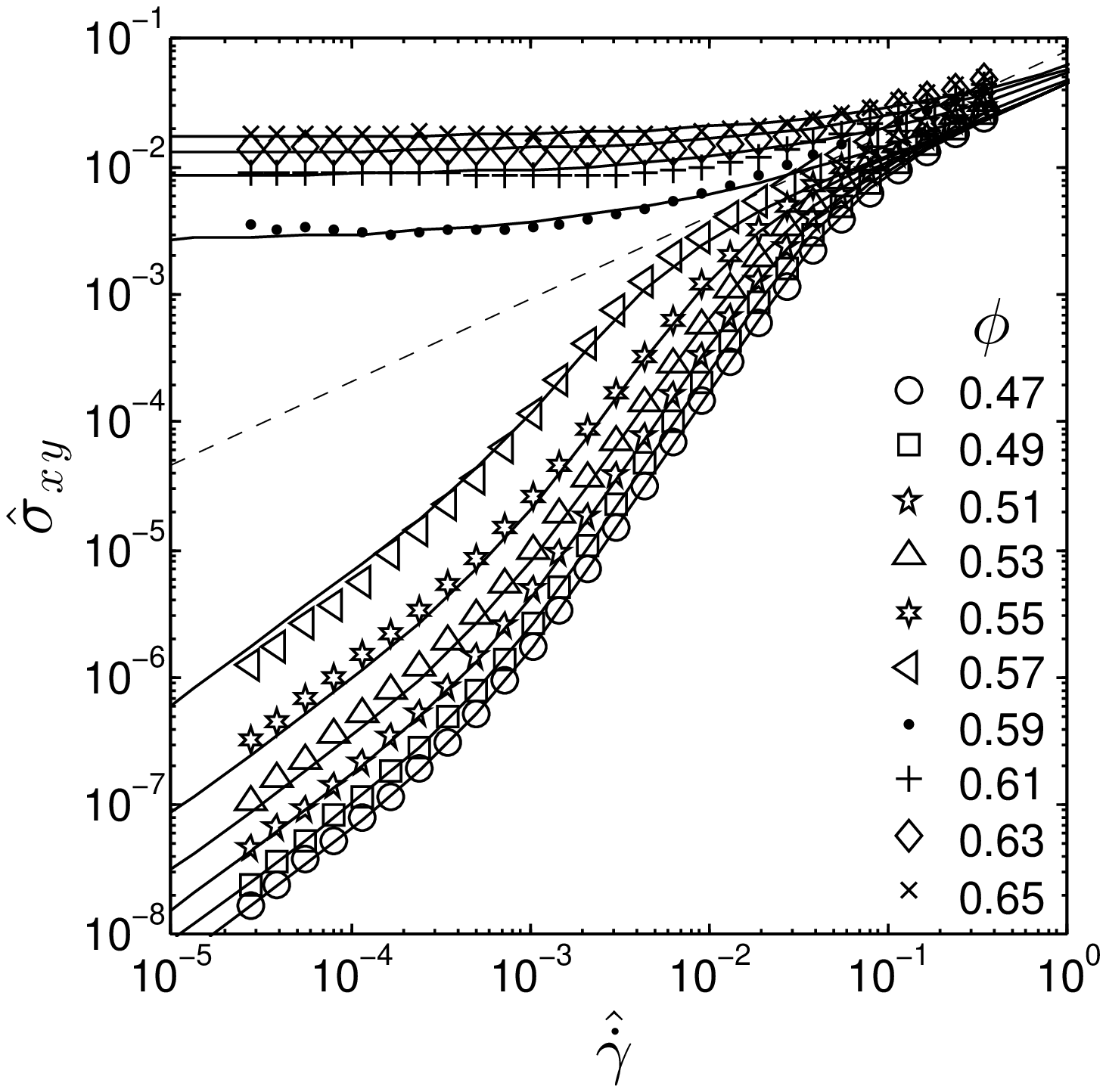}}\hspace{10mm}
      \subfigure[]{
  \includegraphics[height=65mm]{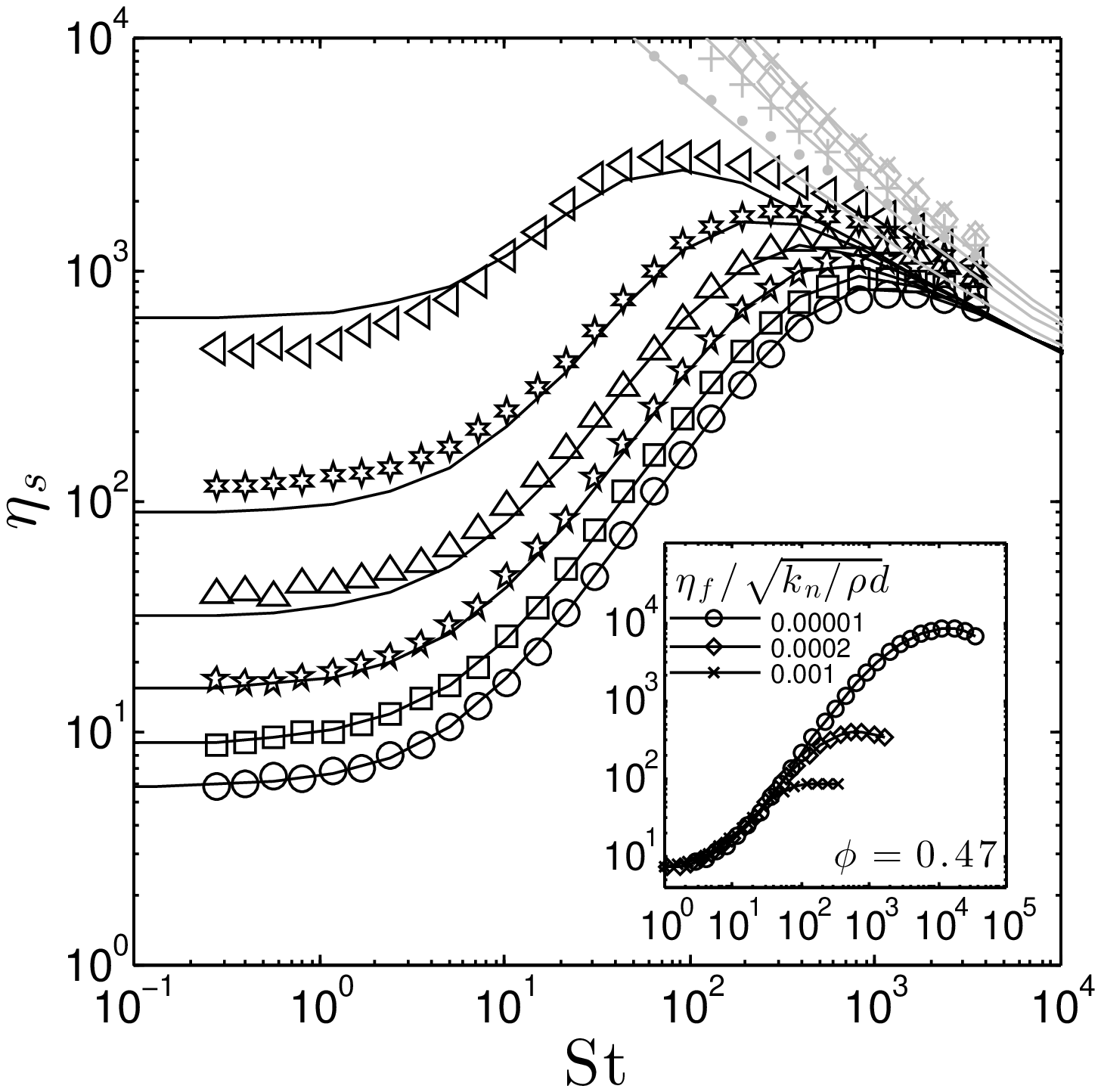}}
     \caption{Flow regime maps for sheared suspension with $\hat{\eta}_f = 2.15 \times 10^{-3}$, at volume fractions given by legend.
(a) Scaled shear stress versus scaled shear rate. Dotted line indicates the critical flow behavior, distinguishing rheology below and above the critical volume fraction $\phi_c$;
(b) Relative suspension viscosity versus Stokes number; Inset: varying $\hat{\eta}_f$ at $\phi = 0.47$, with $\text{St}$ and $\eta_s$ axes.
Symbols represent DEM simulation results, solid lines are predictions from the constitutive model proposed in Section~\ref{sec:const}.
 }
  \label{fig:flowcurve}
\end{figure*}

To achieve homogenous simple shear flow, an assembly of spheres in a 3-dimensional periodic domain is deformed at a constant shear rate $\dot{\gamma}$. Bidispersity at a diameter ratio of $1:1.4$ and volume ratio of about $1:1$ is used to minimize crystallization during flow. A sample assembly under shear is shown in Figure~\ref{fig:particles}, in which the particles are colored into bands in the flow ($x$) direction according to their initial positions as shown in (a) and move to new positions at a later time $t=t_1$ shown in (b), conforming to the simple shear velocity profile. At a shear strain of 0.5, the deformed assembly of particles is mapped to a symmetric position with a strain of $-0.5$. This deformation pattern is repeated ad infinitum to reach a steady state. In practice, we find that a strain magnitude of 1 to 10 is sufficient to reach steady flow, dependent on the shear rate. Simulations at different shear rates have been performed for a range of fixed volume fractions $\phi$ (spanning the jamming transition) and fluid viscosities to probe the bulk rheology and microstructures. It is determined that an assembly of approximately 2000 bidisperse spheres is sufficiently large to capture the bulk rheology independently of the domain size. A larger domain is required in order to capture microstructural phenomena including correlation lengths within the material, and therefore assemblies of approximately 30,000 bidisperse spheres are used for the work presented in Section~\ref{sec:micro}.

The bulk stress is calculated from the particle force and velocity data. It is decomposed into contributions due to the hydrodynamic interaction, the particle-particle interaction and the velocity fluctuation, given by Eqs.~\ref{eq:stressF},~\ref{eq:stressC} and~\ref{eq:stressV}, respectively,  

\begin{subequations}
\begin{align}
\sigma^F &= \frac{1}{V} \sum_i \sum_{i \neq j} \mathbf{r}_{ij} \mathbf{F}^l_{ij} \text{,} 
\label{eq:stressF} \\
\sigma^C &= \frac{1}{V} \sum_i \sum_{i \neq j} \mathbf{r}_{ij} \mathbf{F}^c_{ij} \text{,}
\label{eq:stressC}
 \\
\sigma^V &= \frac{1}{V} \sum_i m_i \mathbf{v}'_i\mathbf{v}'_i \text{,} 
\label{eq:stressV}
\end{align}
\end{subequations}
where $\mathbf{v}'_i$ is the particle velocity after the mean streaming velocity has been subtracted. Data from 20 realizations with randomized initial particle positions are used to obtain ensemble-averaged stresses, which are further averaged over time in the steady-state, as presented in the next section. Under simple shear flow, the relevant stresses that will be discussed are the $xy$ components from each contribution $i$, $\sigma^i_{xy}$, and the mean normal stress (i.e. the pressure) from each contribution $\text{P}^i = \frac{1}{3} (\sigma^i_{xx} + \sigma^i_{yy} + \sigma^i_{zz})$.
The bulk shear and normal stresses, $\sigma_{xy}$ and $\text{P}$, can be obtained by summing the contributions $\sigma^i_{xy}$ and $\text{P}^i$ respectively \cite{Gallier2014}. We note that $\sigma^V_{xy}$ is typically significantly smaller than the other contributions.


\begin{figure*}
  \centering
      \subfigure[]{
  \includegraphics[height=50mm]{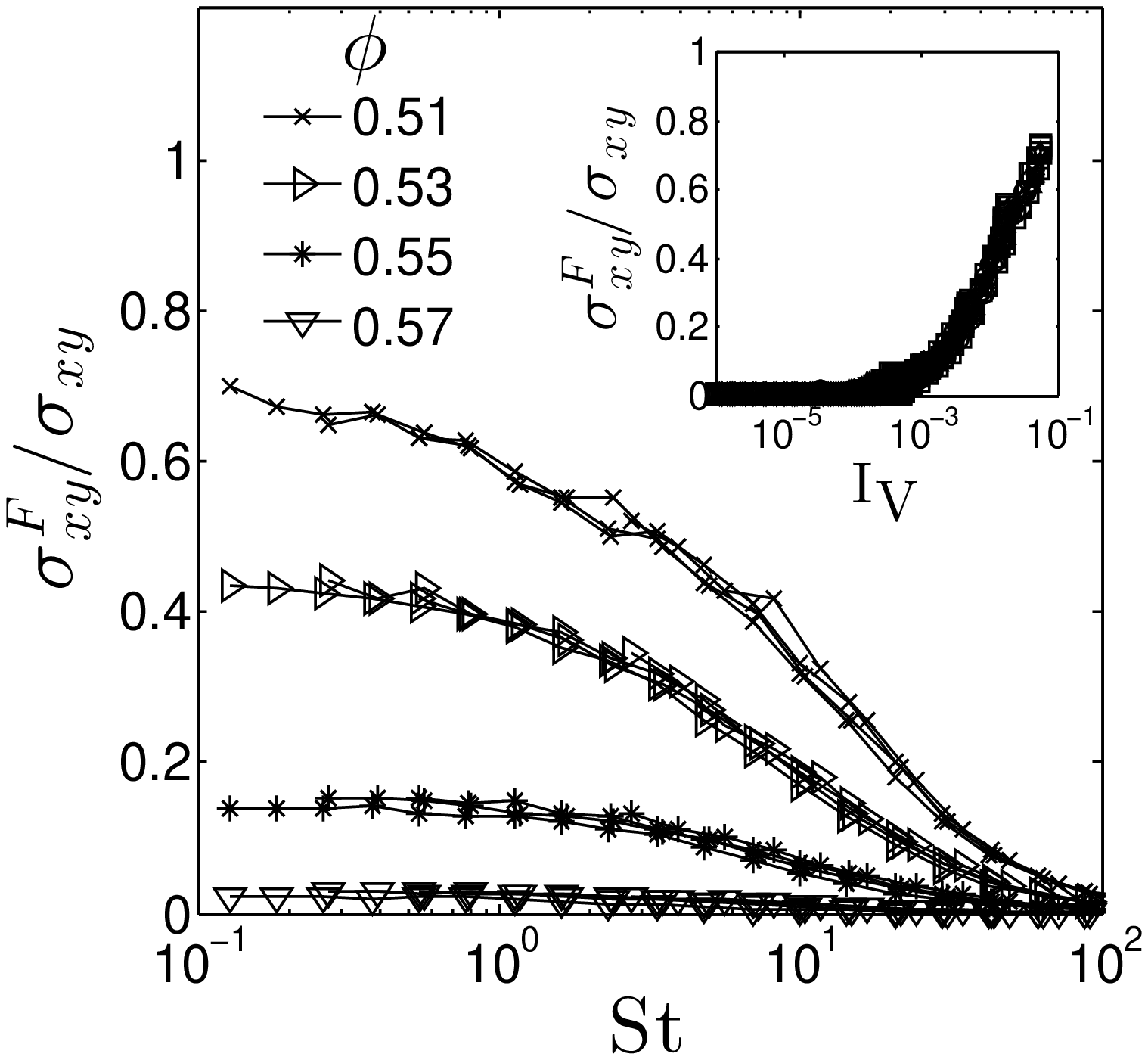}}
      \subfigure[]{
  \includegraphics[height=50mm]{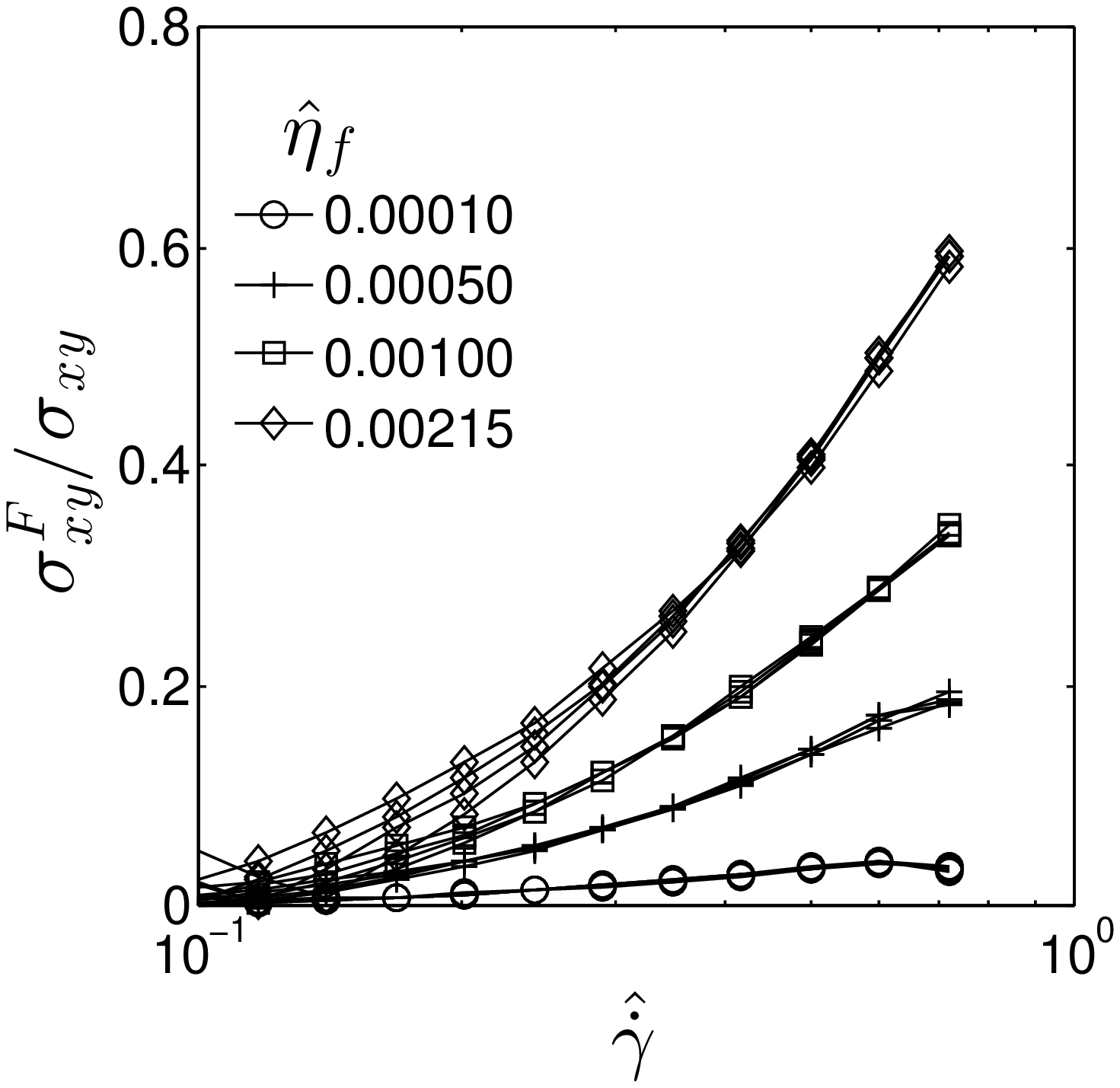}}
        \subfigure[]{
  \includegraphics[height=50mm]{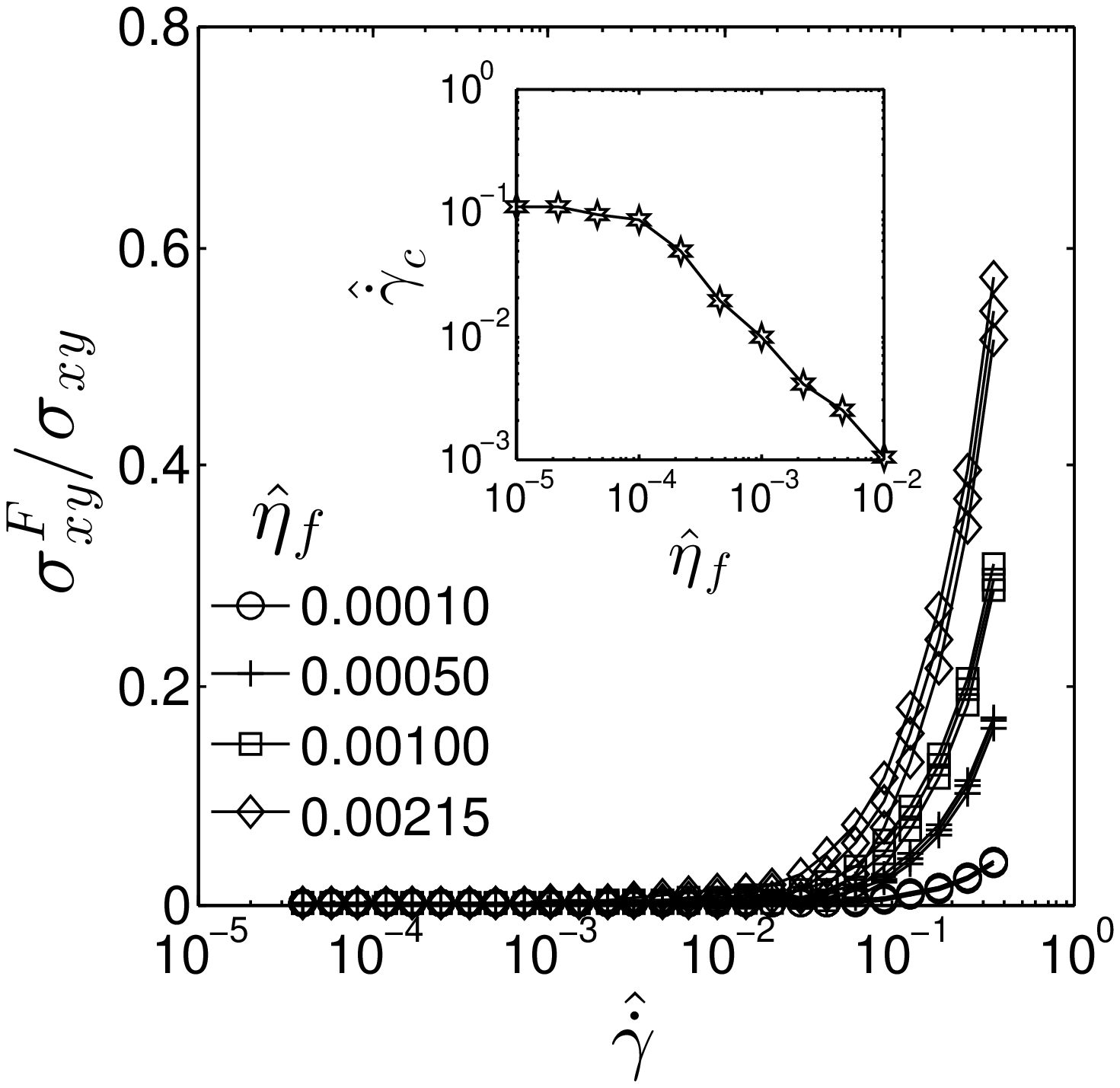}}
     \caption{The varying contribution of fluid stress $\sigma^F_{xy}$ to the total suspension stress $\sigma_{xy}$.
     (a) The viscous-to-inertial transition corresponds to diminishing stress contributions from $\sigma^F_{xy}$, which collapse with $\text{St}$ to $\phi$-dependent branches. Different sets of data correspond to varying $\hat{\eta}_f$; Inset: all the data collapse as a function of $\text{I}_\text{V}$;
     (b) The inertial-to-soft particle transition corresponds to a re-emergence of $\sigma^F_{xy}$ contributions, which collapse with $\hat{\dot{\gamma}}$, to $\hat{\eta}_f$ dependent (and $\phi$-independent) branches. Different sets of data correspond to varying volume fraction below $\phi_c$.
     (c) The quasistatic-to-rate-dependent transition similarly corresponds to a changing stress contribution that collapses with $\hat{\dot{\gamma}}$. Different sets of data correspond to varying volume fraction above $\phi_c$; Inset: critical shear rate for the transition out of the quasistatic state.
     }
  \label{fig:contributions}
\end{figure*}

\section{Bulk rheology \label{sec:bulk}}

A number of relevant timescales must be reconciled in order to fully characterise the data. In the absence of fluid, the shearing timescale, set by $\dot{\gamma}$, competes only with the relaxation time of interparticle contacts, set by $k_n$, so the full rheology of the material at a given volume fraction can be explored by varying the relation between these two, for example using $\hat{\dot{\gamma}} = \dot{\gamma}d/\sqrt{k_n/\rho d}$ \cite{Chialvo2012}, where $\hat{\dot{\gamma}}$ quantifies the departure from hard sphere rheology. The resulting shear stress can be similarly scaled according to $\hat{\sigma}_{xy} = \sigma_{xy}d/k_n$. Such a scaling will be shown to be important to the present simulations at high shear rates (meaning $\hat{\dot{\gamma}} \to 1$) where particle deformations become appreciable relative to the applied flow.
The fluid gives a further relevant timescale, namely a dissipative one set by the fluid viscosity $\eta_f$, which can be related to the shearing timescale by the Stokes number $\text{St} = \rho \dot{\gamma}d^2/\eta_f$. In this case, the shear stress can be appropriately scaled with the viscous stress to give a relative suspension viscosity $\eta_s = \sigma_{xy}/\dot{\gamma}\eta_f$.
A material parameter, given by $\hat{\eta}_f = \eta_f/\sqrt{k_n\rho d}$, is defined to characterise the separation of the viscous dissipation and contact relaxation timescales. The role of this parameter will be demonstrated.
The flow curves obtained from simple shear flow simulations are presented in terms of the $k_n$ and $\eta_f$ scalings, in Figures~\ref{fig:flowcurve}a and \ref{fig:flowcurve}b respectively.

The critical volume fraction $\phi_c$ is identified from Figure~\ref{fig:flowcurve}a as being between 0.57 and 0.59. We first consider the rheology below $\phi_c$.
A quasi-Newtonian regime, in which the suspension viscosity is independent of shear rate (but strongly dependent on $\phi$), emerges at $\text{St}<1$. This regime, in which ${\hat{\sigma}}_{xy} \propto \hat{\dot{\gamma}}$, has been observed experimentally for wet granular materials~\cite{Petekidis2003,Coussot1995a,Fall2010}. 
As the shear rate is increased, the flow behavior transitions from quasi-Newtonian to continuously shear thickening at a Stokes number of approximately 1.  
The $\text{St}>1$ regime, best illustrated for the range $10<\text{St}<100$ in Figure \ref{fig:flowcurve}b, is reminiscent of inertial Bagnoldian flow associated with dry granular materials, where $\hat{\sigma}_{xy} \propto \hat{\dot{\gamma}}^2$. The rheology is described as continuously shear thickening in the inertial regime, in the sense that ${\eta}_s$ scales linearly with ${\dot{\gamma}}$.

The Newtonian-to-Bagnoldian (i.e.~viscous-to-inertial) transition below $\phi_c$ is correlated with the decreasing magnitude of $\sigma^F_{xy}$ relative to $\sigma_{xy}$ as the Stokes number is increased \cite{Lemaitre2009} and particle inertia becomes important, as demonstrated in Figure~\ref{fig:contributions}a. The contribution from $\sigma^F_{xy}$ is roughly independent of shear rate below $\text{St} = 1$, but steadily drops off for $\text{St} > 1$, indicating that the fluid plays a diminishing role in the suspension rheology as the particle inertia is increased. It is found that the behaviour of $\sigma^F_{xy} / \sigma_{xy}$ for varying values of material parameter $\hat{\eta}_f$ can be collapsed to volume fraction dependent branches when plotted with the Stokes number, demonstrating our previous assertion that the rheology in the viscous-to-inertial transition is governed only by the Stokes number (and the volume fraction).
This result can be generalized for all $\phi$, ${\dot{\gamma}}$ and ${\eta}_f$ using the viscous number $\text{I}_\text{V}$, as shown in the Inset of Figure~\ref{fig:contributions}a. Above some critical viscous number, the stress arising from fluid effects becomes significant. A similar result was suggested by Huang et al.~\cite{Huang2004}, although they use the Leighton number ($\text{Le} = \eta_f \dot{\gamma} / \sigma_{xy}$) rather than the viscous number.

 This viscous-to-inertial transition can also be reconciled at the microscopic level by examining the relative magnitude of $\textbf{F}^l_{ij}$ and $\textbf{F}^c_{ij}$ for interacting pairs, and by appealing to the Sommerfield number associated with lubrication theory $s = \eta_f v/ d \sigma_{xy}$, where $v$ represents some relative velocity between the particle surfaces, dependent on the bulk shear rate \cite{Fernandez2013}. Below a critical $s$, the lubrication films between particles rupture and mechanical contacts are initiated, at which point the stress response becomes contact-dominated rather than fluid-dominated. The critical Stokes number of 1 for the onset of the viscous-to-inertial transition relates to a critical Sommerfield number at which lubrication films begin to break down, leading to the reduction of the $\sigma^F_{xy}$ magnitude.

As the shear rate is increased further ($\hat{\dot{\gamma}}\to 1$ and $\text{St}>100$ in Figures \ref{fig:flowcurve}a and b respectively), the flow exits the continuously shear thickening inertial regime, and a soft particle rheology is realised. The nature and location of this high shear rate asymptotic flow behaviour, in which the dependence on $\phi$ becomes small, is highly dependent on the material parameter $\hat{\eta}_f$. As discussed previously, the viscous-to-inertial transition occurs around $\text{St} = 1$, and we assert here that the transition to high shear rate, soft particle rheology occurs as $\hat{\dot{\gamma}} \to 1$. The material parameter $\hat{\eta}_f$ sets the separation between the two dimensional shear rates $\dot{\gamma}$ associated with the St and $\hat{\dot{\gamma}}$ limits, effectively determining the range of shear rates over which inertial flow can be observed. For example, a suspension of very hard particles, for which $k_n \to \infty$ and $\hat{\eta}_f \to 0$, may transition between viscous and inertial flow regimes at experimentally accessible shear rates, while it would be impossible to access their ``soft particle'' rheology as $\hat{\dot{\gamma}} \to 1$, which would correspond to $\text{St} \to \infty$. Conversely, for soft particles, $\text{St} = 1$ may occur very close to $\hat{\dot{\gamma}} = 1$. Our data therefore explore the crossover from hard sphere rheology at low $\hat{\eta}_f$ in the vicinity of $\text{St} =1$ to soft sphere rheology at high $\hat{\eta}_f$ as the condition of $\text{St} = 1$ approaches  that of $\hat{\dot{\gamma}} = 1$. The role of $\hat{\eta}_f$ is demonstrated in the Inset of Figure \ref{fig:flowcurve}b, which illustrates that inertial flow is observed over a wider range of $\text{St}$ as $\hat{\eta}_f \to 0$.

Interestingly, the rheology (not just the location) of the soft particle limit is dependent on the value of $\hat{\eta}_f$.
 In the low $\hat{\eta}_f$ limit ($\hat{\eta}_f = 5 \times 10^{-5}$ data shown in Inset of Figure \ref{fig:flowcurve}b), shear thinning, ``intermediate'' \cite{Chialvo2012}, rheology is observed as $\hat{\dot{\gamma}} \to 1$, with $\hat{\sigma}_{xy}~\propto~\hat{\dot{\gamma}}^{0.5}$, consistent with previous experiments in soft, highly deformable particles~\cite{Nordstrom2010a}. The origin of this shear thinning scaling exponent is still uncertain, though it may relate to the large particle deformations that occur at such high shear rates. A switch back to viscous, Newtonian scaling is observed at $\hat{\dot{\gamma}}\to1$ for high $\hat{\eta}_f$ as demonstrated in the Inset of Figure~\ref{fig:flowcurve}b by a shift back to rate-independent suspension viscosity for $\hat{\eta}_f = 1 \times 10^{-3}$. As with the Newtonian-to-Bagnoldian transition around $\text{St} = 1$, the transition between intermediate and viscous soft particle rheology as $\hat{\dot{\gamma}} \to 1$ can be correlated with an increase in the magnitude of $\sigma^F_{xy}$ relative to $\sigma_{xy}$ as $\hat{\eta}_f$ is increased, Figure \ref{fig:contributions}b. Notably, these data now collapse for different volume fractions onto $\hat{\eta}_f$ dependent branches, when plotted with $\hat{\dot{\gamma}}$, demonstrating the loss of $\phi$ dependence and contrasting with the $\text{St}$ collapse observed for the viscous-to-inertial transition.

Above $\phi_c$, a contact dominated, quasistatic ($\hat{\sigma}_{xy} \propto \hat{\dot{\gamma}}^0$) regime exists for $\hat{\dot{\gamma}} \ll 1$ as demonstrated in Figure~\ref{fig:flowcurve}a. Although the quasistatic regime is inherently a ``soft'' particle phenomena, in the sense that flow above $\phi_c$ without ordered localization is not possible for ideally hard particles, it is not characterised by large particle deformations and volume fraction independence, except perhaps for $\phi \gg \phi_c$. As $\hat{\dot{\gamma}}$ approaches 1, however, the flow transitions to a rate-dependent, soft particle state, coinciding with that of the $\phi < \phi_c$ case, and consistent with an experimentally observed quasistatic-to-viscous transition observed for suspensions of soft polystyrene beads \cite{Huang2004}.
As with the transitions below $\phi_c$, the transition out of the quasistatic regime can be well correlated with the emerging dominance of the fluid contribution relative to the contact contribution to the stress, Figure~\ref{fig:contributions}c. In all cases, the stress in the rate-independent, quasistatic regime (where $\hat{\dot{\gamma}} \ll 1$) is contact dominated, so $\sigma^F_{xy}/\sigma_{xy} \to 0$. As $\hat{\dot{\gamma}} \to 1$, and the rheology becomes rate dependent, the fluid contribution becomes significant.

The nature of the rate-dependent regime depends strongly on $\hat{\eta}_f$, as with the $\phi<\phi_c$ case. We define a critical shear rate $\hat{\dot{\gamma}}_{c}$ at the point where the fluid contribution begins to increase, corresponding to the quasistatic-to-rate-dependent transition. Consistent with recent experimental findings~\cite{Huang2004}, the critical shear rate is found to be a linearly decreasing function of $\hat{\eta}_f$, for sufficiently high $\hat{\eta}_f$, Inset of Figure~\ref{fig:contributions}c. We extend the study to lower fluid viscosities, however, and find that at low $\hat{\eta}_f$, a plateau is observed. This plateau indicates that below a certain critical $\hat{\eta}_{fc}$ (of the order $10^{-4}$ from the Inset of Figure \ref{fig:contributions}) the flow is never $\hat{\eta}_f$-dependent, providing a critical value of the material parameter to distinguish between intermediate $\hat{\sigma}_{xy} \propto \hat{\dot{\gamma}}^{0.5}$ ($\hat{\eta}_f <  \hat{\eta}_{fc}$) and viscous $\hat{\sigma}_{xy} \propto \hat{\dot{\gamma}}$ ($\hat{\eta}_f > \hat{\eta}_{fc}$) rheology in the soft particle limit.

As a summary, we observe regime transitions between viscous, inertial, quasistatic, intermediate and ``soft'' viscous shear flow as the timescales for viscous dissipation ($\text{St}$) and particle contact relaxation ($\hat{\dot{\gamma}}$) are varied. The mechanism for such transitions can be ascribed to the variation of the relative importance of the fluid and contact stress contributions. The bulk flow can therefore be effectively described by considering a background dry granular rheology, governed by the $\hat{\dot{\gamma}}$ scaling, superimposed with a viscous stress, governed by the $\text{St}$ scaling. Such a constitutive model will be presented in Section~\ref{sec:const} after we further examine the microstructure of the flow in the next section.

\section{Microstructure \label{sec:micro}}

Particle level dynamics and structures are studied in order to shed light on the microscale phenomena responsible for the bulk rheology in each flow regime.

\begin{figure}
      \subfigure[Inertial]{
  \includegraphics[height=40mm]{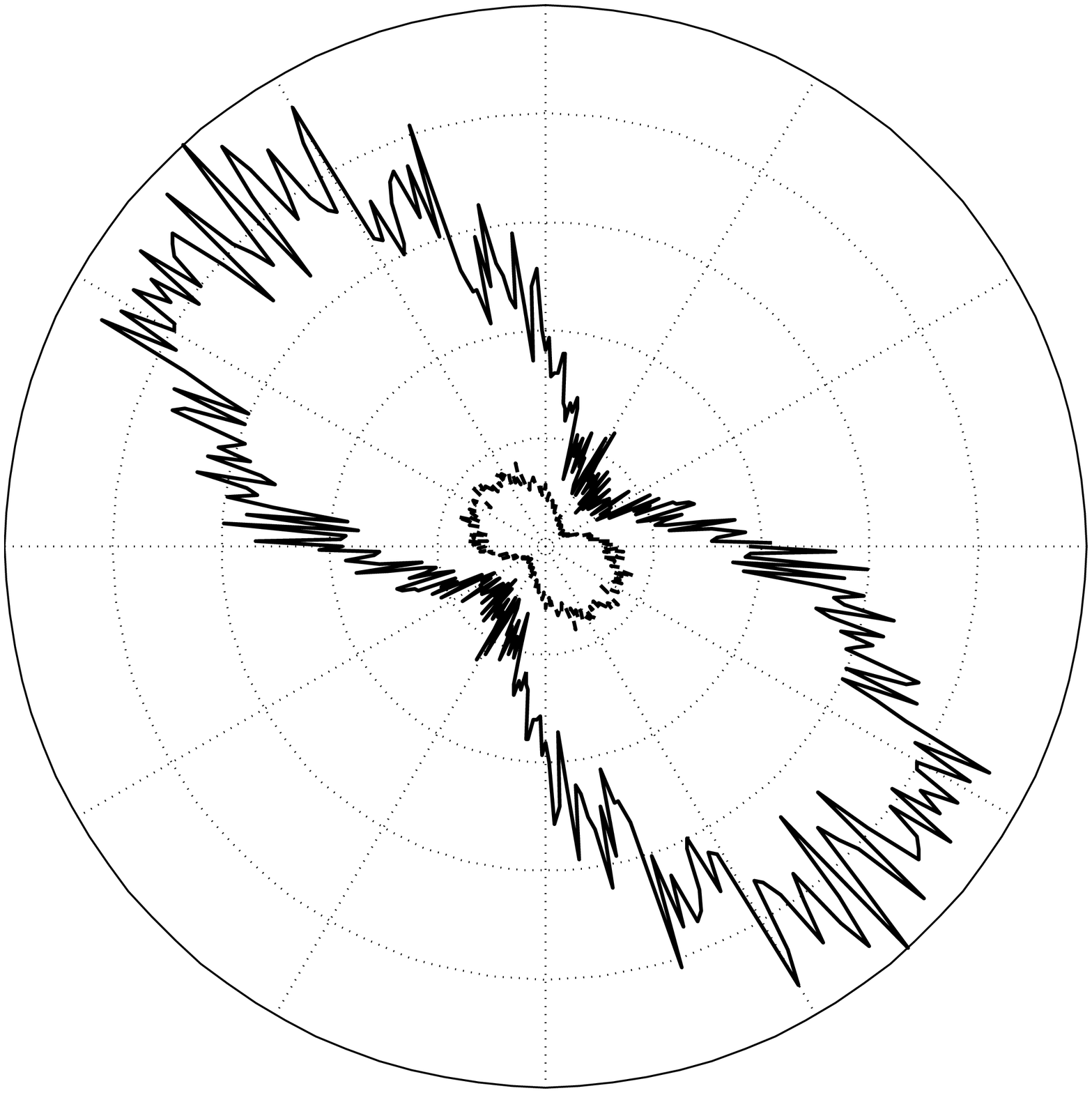}}
      \subfigure[Viscous]{
  \includegraphics[height=40mm]{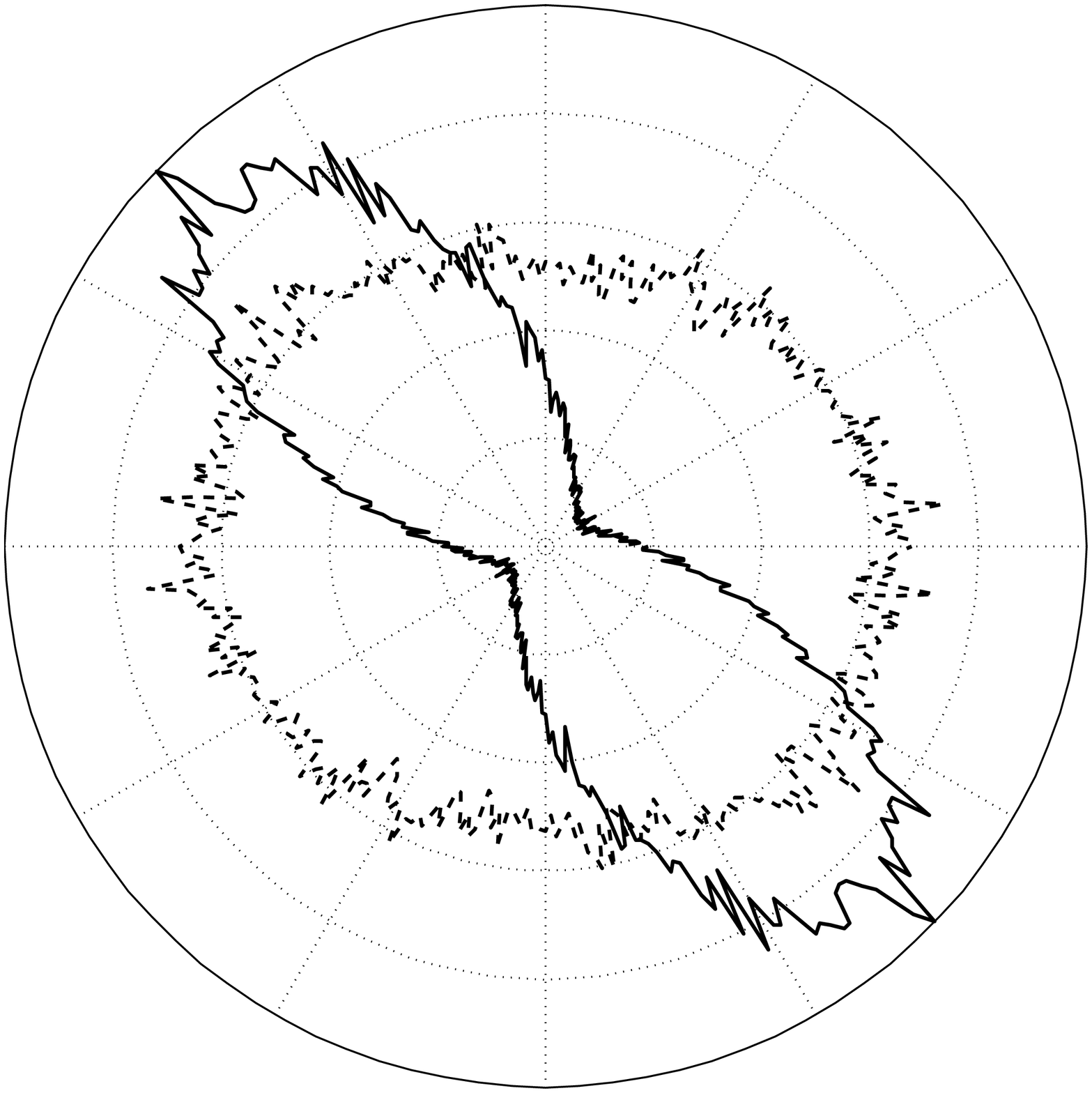}}
        \subfigure[Quasistatic]{
  \includegraphics[height=40mm]{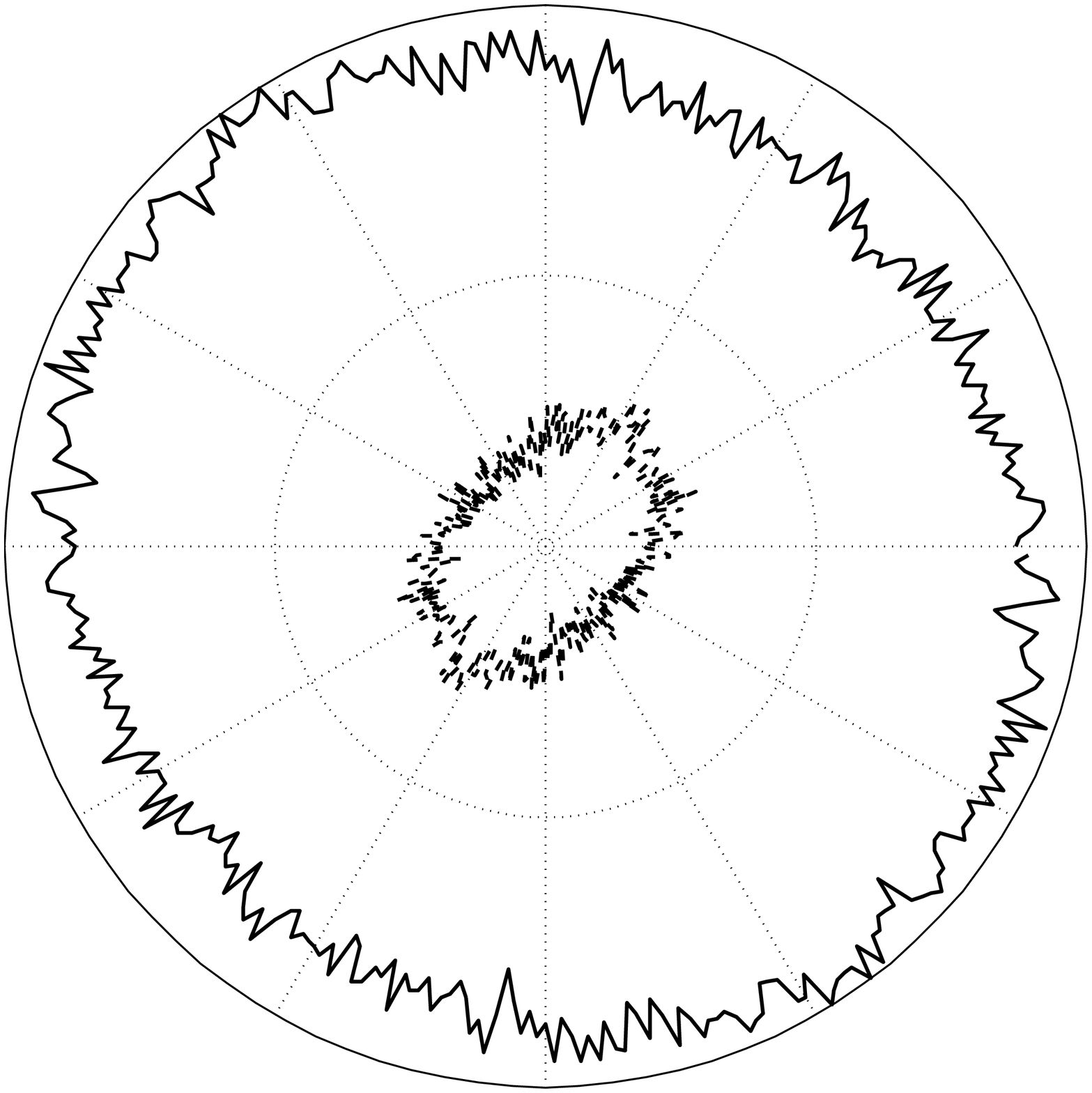}}
          \subfigure[Intermediate]{
  \includegraphics[height=40mm]{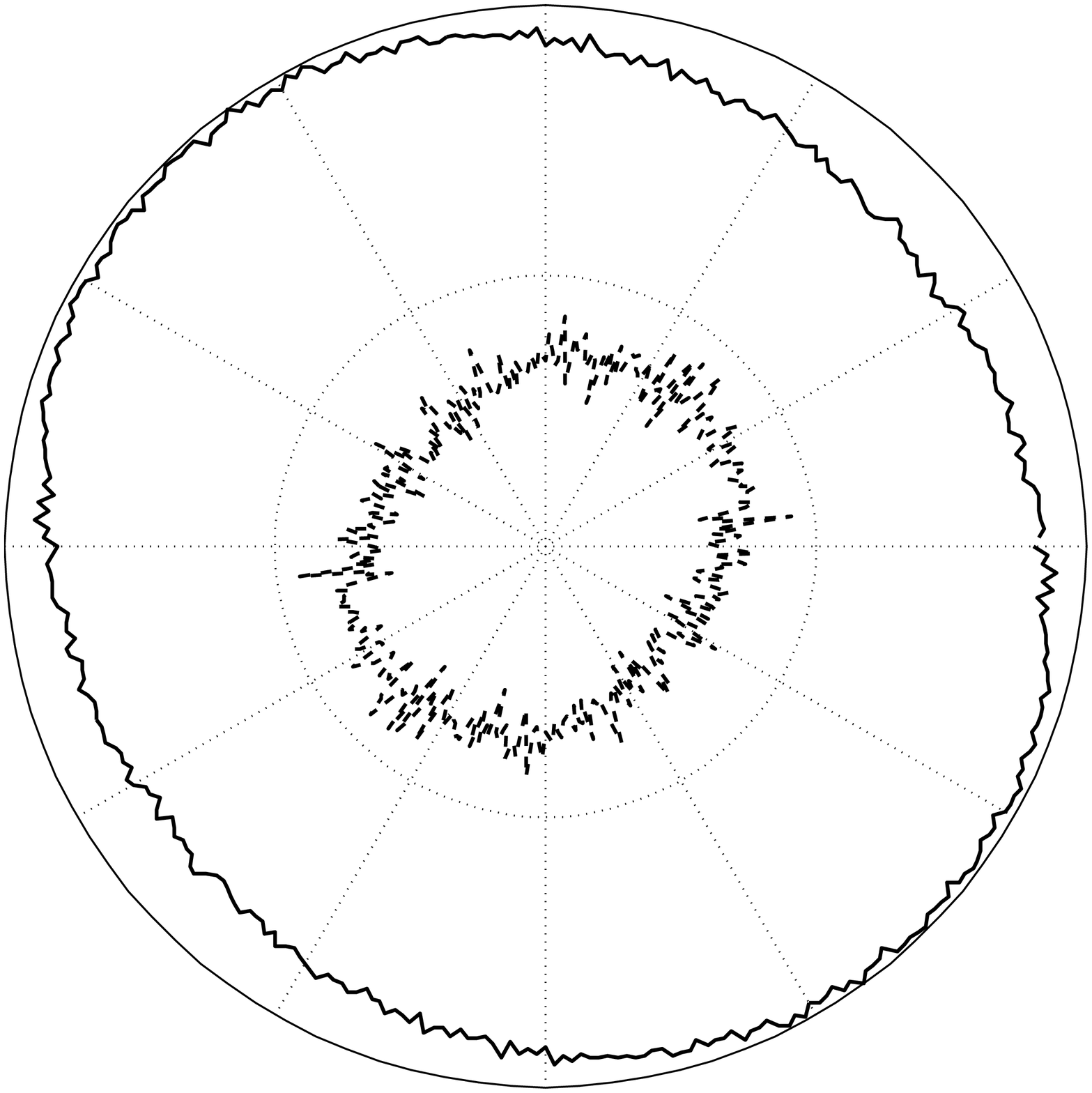}}
        \subfigure[Soft viscous]{
  \includegraphics[height=40mm]{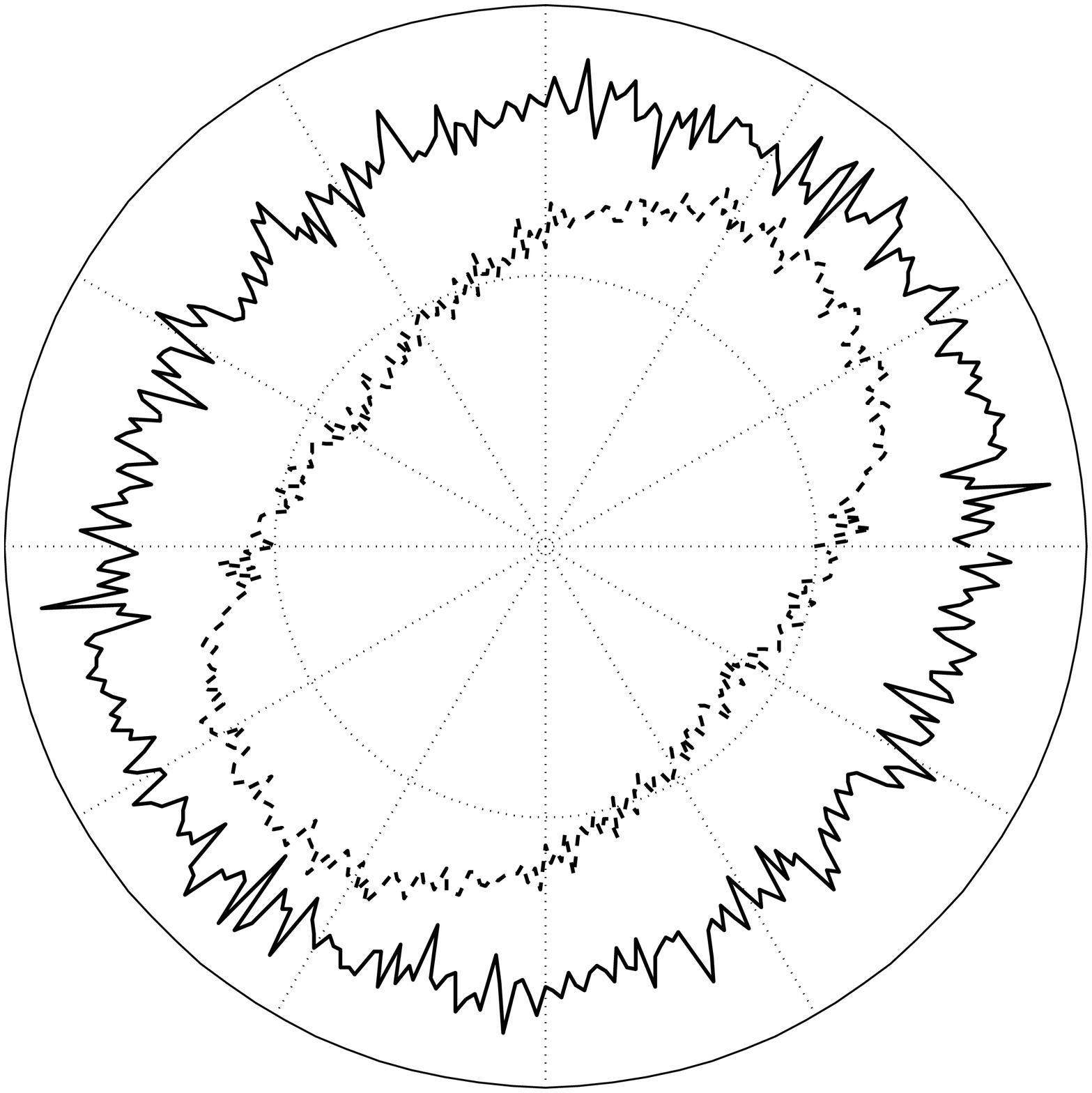}}
      \subfigure[Velocities]{
  \includegraphics[height=40mm]{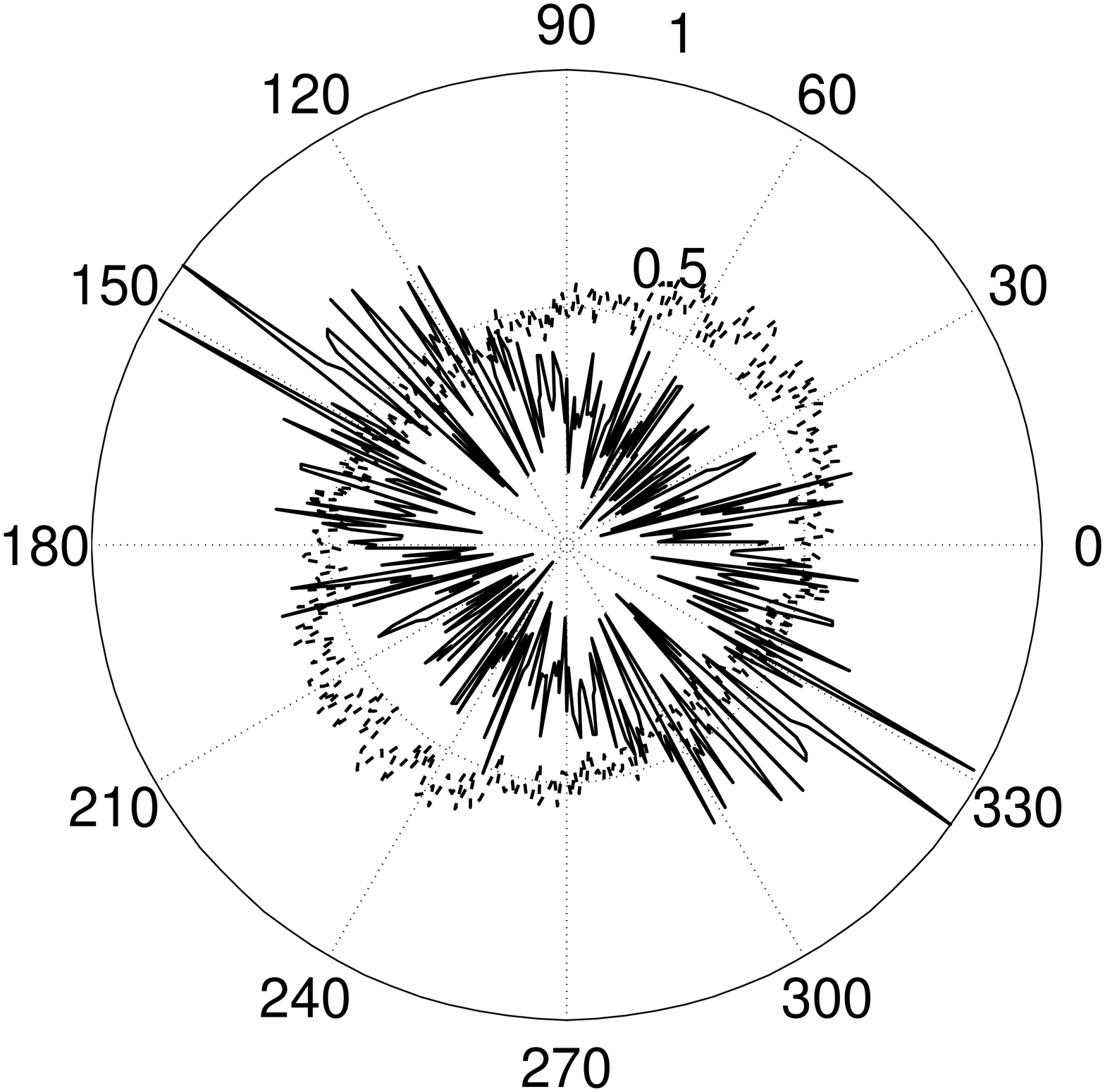}}
     \caption{(a) - (e) Radial force distributions for contact (solid line) and lubrication (dashed line) forces, for each of the flow regimes described in Section \ref{sec:bulk}. The statistics include all particle pairs over all time steps in steady state and the forces are scaled in magnitude by the respective maximal values. (f) Radial distribution of the relative velocity between interacting particles for $\phi=0.5$ (solid line) and $\phi = 0.65$ (dashed line). 
	 \label{fig:radial}
 }
\end{figure}

 The distributions of pairwise particle forces (defined in Equations \ref{eq:lube} and \ref{eq:cont}) are plotted in Figures~\ref{fig:radial}a-e for each of the flow regimes identified previously, illustrating the radial directions (in the $xy$-plane) and relative magnitudes of the lubrication and particle contact forces.

In the inertial regime (Figure~\ref{fig:radial}a), there is a clear alignment of forces along the direction of compression, as expected for a collisional flow \cite{DaCruz2005}, demonstrating that forces are transmitted through the material along a principal axis opposing the shear flow. The contact forces dominate over the lubrication forces, consistent with observations of the bulk contact stress dominance in this regime. 
As the Stokes number is reduced and the role of the lubrication stress becomes dominant, the force directions are somewhat different. In the viscous regime (Figure~\ref{fig:radial}b), the magnitude of the lubrication forces becomes comparable to that of the contacts and their orientation becomes significantly more isotropic. This suggests that the viscous lubrication films are arresting much of the rapid, inertial particle motion by suspending the particles in a more uniform, isotropic and interconnected fluid film network. Isolating the particle contact forces, we find that the contact network retains its strong alignment even though the resulting observed rheology is dominated by the stress coming from lubrication forces.
In the quasistatic regime (Figure~\ref{fig:radial}c), the contact forces become significantly more isotropic. This is attributed to the jammed state being a more interconnected network, where forces are transmitted not via collisions along a shearing direction, but through persistent contacts that compose mesoscale chains and clusters. While the contact forces appear to be completely isotropic, or perhaps aligned very slightly in the compressive axis, there is a small alignment of the lubrication force along the extensional axis, indicating that the lubrication forces are acting in an opposing direction to the contact forces. It is noted that the contact forces are significantly greater that the lubrication forces in this regime, consistent with the observations from bulk stresses that quasistatic rheology is dominated by particle contacts.
The radial forces in the intermediate regime (Figure~\ref{fig:radial}d) are consistent with those for quasistatic flow, though the role of the lubrication force becomes slightly greater, in agreement with the bulk stress contributions presented in Figure \ref{fig:contributions}c for increasing $\hat{\dot{\gamma}}$ at low $\hat{\eta}_f$.
 In the soft viscous regime (Figure~\ref{fig:radial}e), the dominance of the lubrication force increases further, again consistent with Figure \ref{fig:contributions}c for increasing $\hat{\eta}_f$. 
It is also observed that the contact forces become aligned along the extensional direction, i.e. with the direction of shear, in this regime, similar to the lubrication forces.

It is further found, independently of Stokes number, that the relative velocities of interacting particles (through both lubrication and mechanical contact) are generally aligned along the compressional axis for $\phi<\phi_c$, as shown by the solid line in Figure~\ref{fig:radial}f. The relative velocities are calculated by first subtracting the mean streaming velocity from all particles, then determining the relative magnitude and direction of the velocities between neighbouring particles (i.e. $\mathbf{v}_\text{rel} = \mathbf{v}'_i - \mathbf{v}'_j$). For $\phi>\phi_c$, however, a significant number of relative particle velocity vectors now align with the extensional axis (dashed line in Figure~\ref{fig:radial}f), the opposite to what was observed for $\phi<\phi_c$. For the quasistatic and intermediate regimes, this results in the changing alignment of the lubrication forces, but since the contact forces are dominant, there is no change in alignment of the net force. For the soft viscous regime, however, the velocity-dependent lubrication force $\textbf{F}^l_{ij}$ becomes significant, and a change in both the contact and lubrication force orientations is observed. 

A great wealth of information regarding the dynamics of suspension flow may be obtained from such radial plots, for example the relationships between force orientation and bulk stresses considering both compressive and tensile stresses and the precise role of the interstitial fluid, however we limit the present work to correlating the above microscopic signatures with the observed bulk rheological regimes.

\begin{figure}
      \subfigure[]{
  \includegraphics[height=38mm]{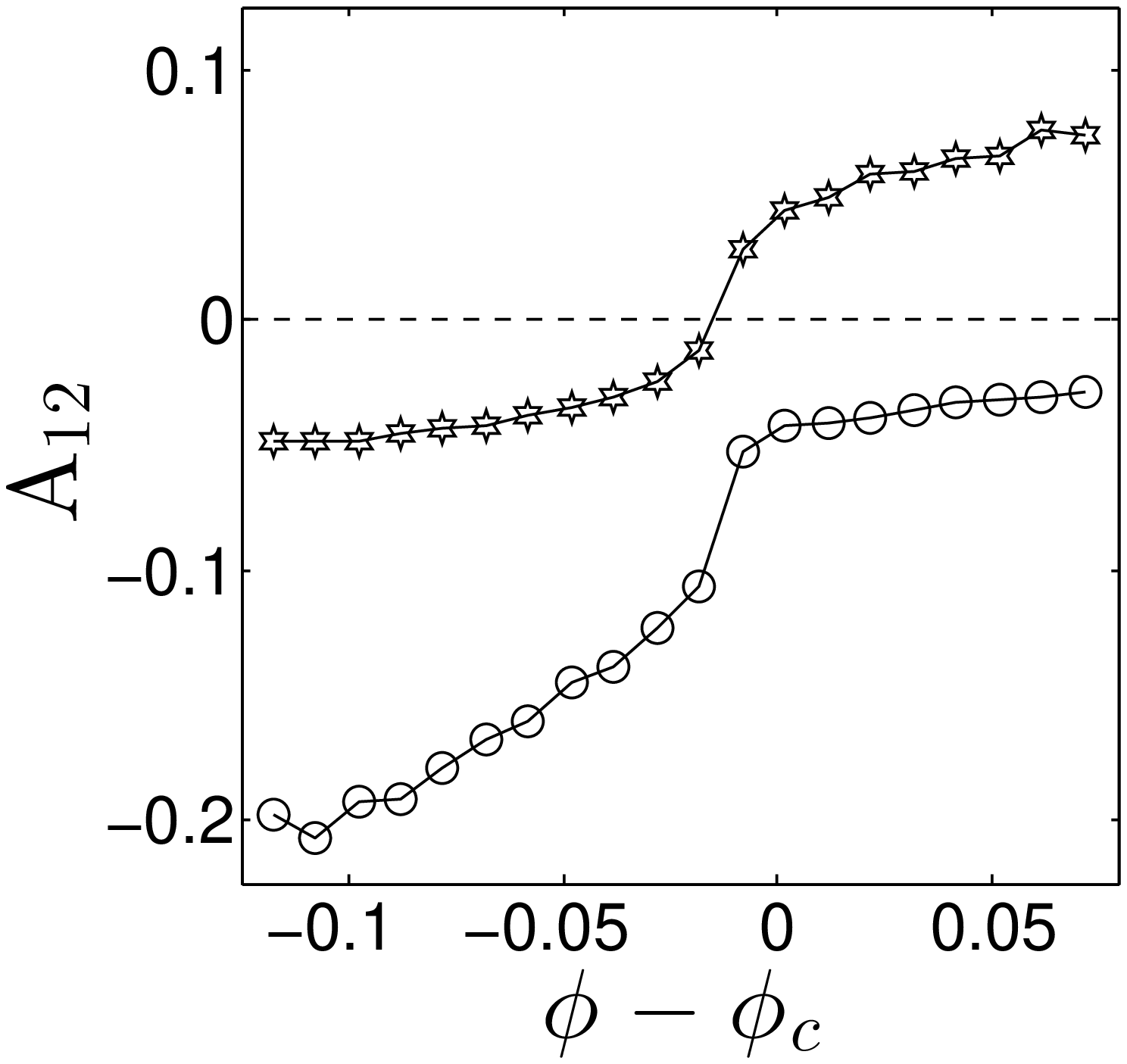}}
      \subfigure[]{
  \includegraphics[height=38mm]{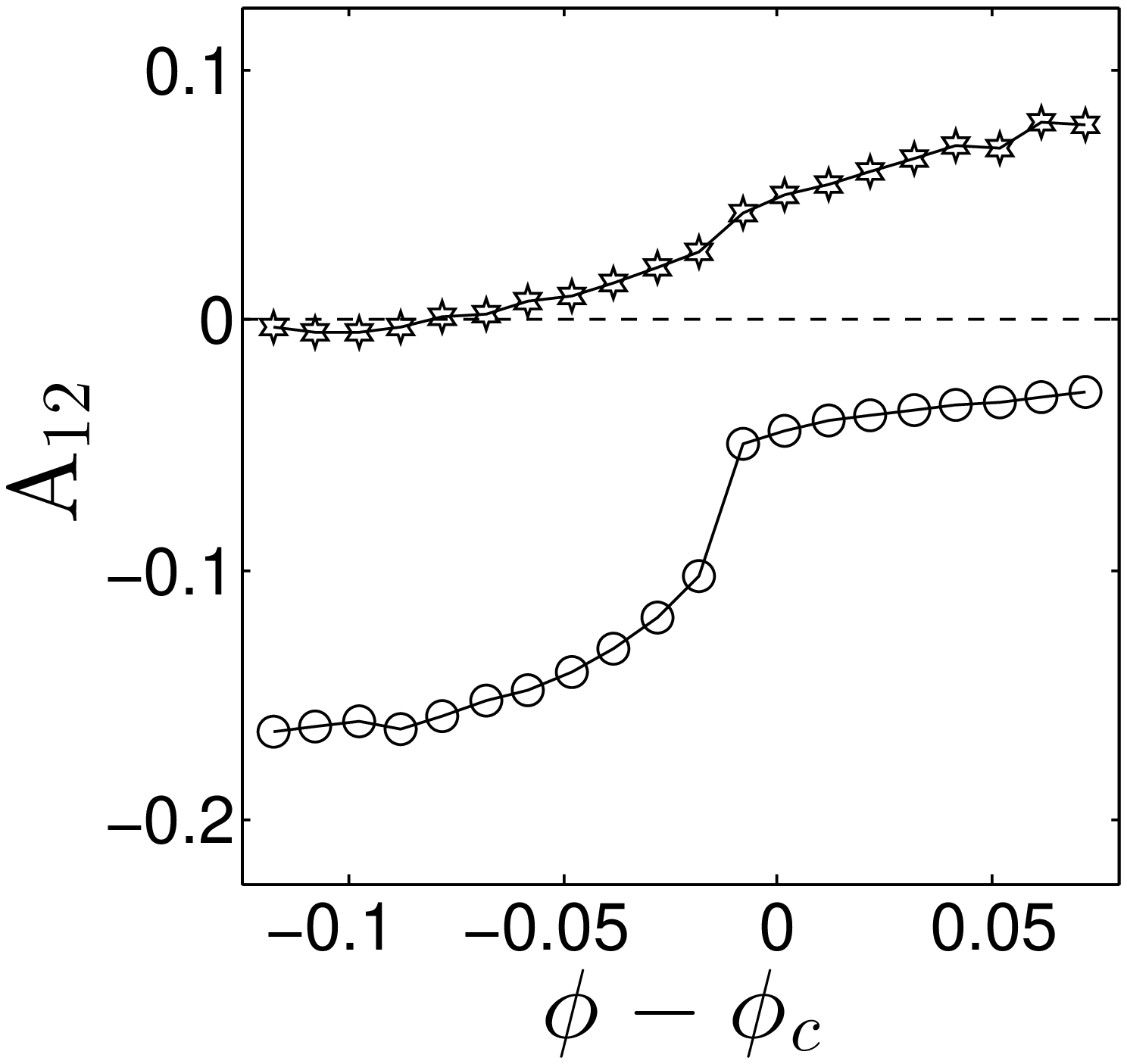}}
  \caption{Shear component of the fabric tensor $\text{A}_{12}$ plotted against volume fraction at (a) $\text{St} = 10$ and (b) $\text{St} = 0.1$. Open circles show the contact network; stars show the fluid lubrication network.
  \label{fig:fabric}
     }
\end{figure}

To corroborate the above observations of the transitions in the microscopic dynamics, a fabric tensor characterising the contact network microstructure~\cite{Sun2011} is constructed for each particle assembly according to
\begin{equation}
\mathbf{A} = \frac{1}{\text{N}_c} \sum^{\text{N}_c}_{\alpha=1} \mathbf{n}_{ij} \mathbf{n}_{ij} - \frac{1}{3} \mathbf{I} \text{,}
\end{equation}
where $\text{N}_c$ is the number of pairwise contacts and $\mathbf{I}$ is the identity tensor. Two variants of the fabric tensor are computed. In the first, we sum over all particles that are in mechanical contact ($h<0$), while in the second we sum over all particles that are separated by a lubrication film (those particles that are separated by a fluid layer of thickness $h$, where $0< h\leq0.05d_{ij}$), giving two separate quantifications of the microstructure, pertaining to contact and lubrication forces respectively. The extent of the structural anisotropy can be quantified using the $xy$-component of the fabric tensor, $\text{A}_{12}$. Alignments that oppose the shear flow of the material are negative in $\text{A}_{12}$, while completely isotropic force networks will give $\text{A}_{12} = 0$. The variation of $\text{A}_{12}$ across volume fractions at $\text{St} = 10$ and $\text{St} = 0.1$ is given in Figures~\ref{fig:fabric}a and b, respectively.

At $\phi < \phi_c$ and $\text{St} = 10$ (Figure~\ref{fig:fabric}a), $\text{A}_{12}$ is negative for both the contact and lubrication network, in agreement with the significant alignment of both forces shown in the radial force plot in Figure~\ref{fig:radial}a. As $\phi$ is increased above $\phi_c$, $\text{A}_{12}$ for the contact network tends to nearly zero (approximately $-0.04$, consistent with previous work in dry, quasistatic shear flows~\cite{Sun2011}), while the lubrication network becomes aligned in the opposite direction. The net effect of this transition is a shift from strongly aligned to almost isotropic net force distribution (Figures~\ref{fig:radial}a and \ref{fig:radial}c), since the opposite alignment occurring in the lubrication network has only a minor contribution to the total force in the quasistatic regime. 
At $\text{St} = 0.1$ (Figure~\ref{fig:fabric}b), the same behavior is observed in the contact network across $\phi_c$, consistent with the radial contact force plots in Figure \ref{fig:radial}a and b. The fabric of the lubrication network is, however, different. Below $\phi_c$, the lubrication fabric tends to zero in this case, corresponding to the isotropic state of the lubrication contacts presented in Figure \ref{fig:radial}b. 

Above $\phi_c$, the behaviour of the fabric appears to be independent of the Stokes number.
The contact network tends towards a nearly isotropic state, while the anisotropy of the lubrication network increases in the extensional direction.
 This anisotropy, together with the relative velocities in the same direction, results in the lubrication force alignment along the extensional axis, which becomes dominant in the soft viscous regime.

\begin{figure}
        \subfigure[]{
  \includegraphics[height=40mm]{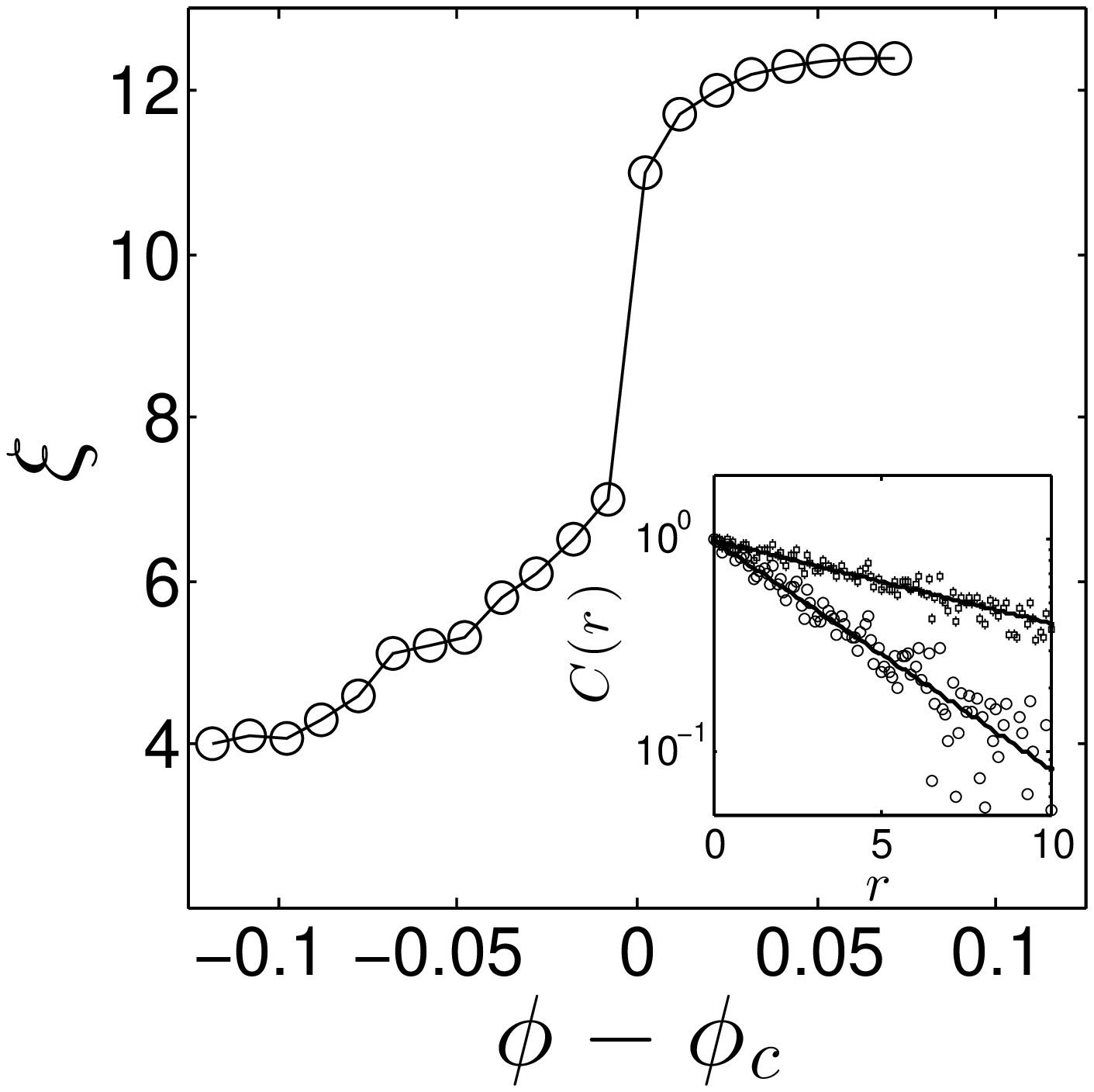}}
      \subfigure[]{
  \includegraphics[height=40mm]{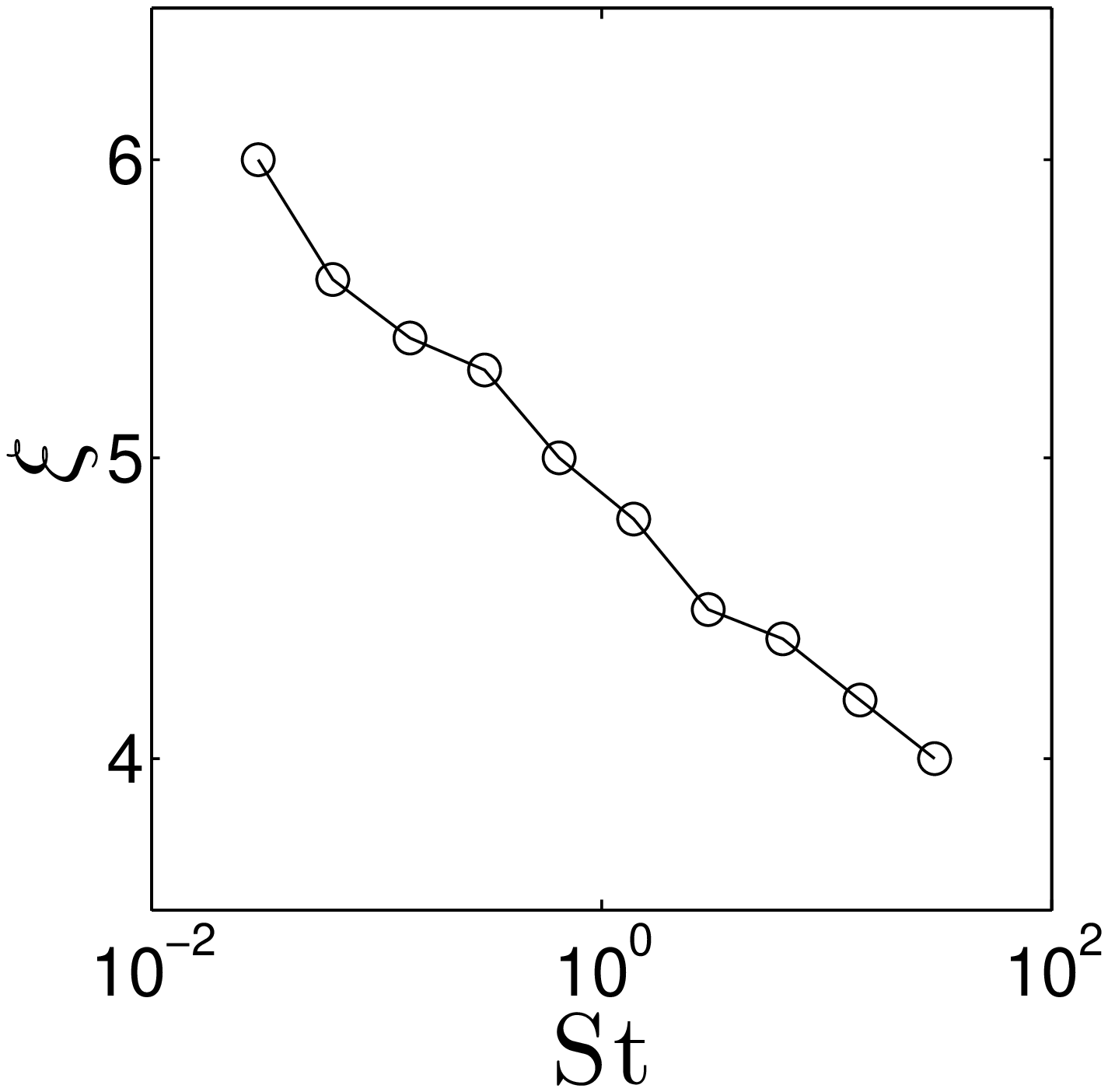}}
  \caption{(a) Velocity correlation length $\xi$ at $\text{St} = 10$ as a function of volume fraction and (b) as a function of $\text{St}$ at $\phi=0.50$ . 
   \label{fig:corr_length}
     }
\end{figure}

The collective motion is further quantified by the velocity correlation length~\cite{Lois2007a,Sun2006}, defined according to
\begin{equation}
c({r}) = \frac{\sum_i \sum_{j>i} \overline{\mathbf{v}}_i \cdot \overline{\mathbf{v}}_j \delta({r}_{ij} - {r})}{\sum_i \sum_{j>i}\delta({r}_{ij} - {r})} \text{,}
\end{equation}
where ${r}_{ij}$ is the center-to-center vector for particles $i$ and $j$ and $\overline{\mathbf{v}}_i$, $\overline{\mathbf{v}}_j$ are particle velocity vectors averaged over a length of time sufficient to give an averaged particle displacement due to the mean flow of approximately $0.5d$.
From this expression we can quantify the extent to which the velocity of a particle is correlated with the velocities of its neighboring particles, on average. It is found that the correlation decays approximately exponentially with the distance ${r}$ (Figure~\ref{fig:corr_length}a, Inset). We fit the data to an exponential function $C({r}) = ke^{-{r}/\xi}$, characterizing the correlation length according to the value of $\xi$~\cite{Lois2007a}. The dependence of this correlation length on volume fraction and Stokes number is given in Figures~\ref{fig:corr_length}a and \ref{fig:corr_length}b, respectively. The correlation length increases with volume fraction, suggesting that as more particles are added, and as gaps between particles reduce in size, there is increasing collective motion in the material. At $\phi_c$ there is a large jump in $\xi$ to about half the domain length, indicating that the correlation extends across the whole domain as the material enters a quasistatic state. The forming of collectively moving clusters of particles is consistent with the radial force plots obtained above $\phi_c$. Below $\phi_c$, correlation lengths are short and forces are dominated by collisions along the compressive axis. Above $\phi_c$, the forces are more uniformly distributed radially, as the particles move more as collective clusters.

 As the Stokes number is increased, the longer range forces arising from lubrication effects become decreasingly dominant, particle inertia plays a greater role, and the correlation length decreases accordingly (Figure~\ref{fig:corr_length}b). Again, this is consistent with the observed radial force distributions. As the Stokes number decreases, the particles become suspended in an increasingly strong network of lubrication films, which retard the inertial, collisional behavior, leading to increased correlation lengths and more isotropic force distributions.

The microstructural details discussed above explain well the mechanisms underlying the transitions in the bulk rheology. When particle collisions dominate, the forces and fabric are anisotropic at the microscale and velocities are correlated over very short lengths. Such dynamics, characteristic of a collisional regime, give rise to the inertial bulk rheological response observed for $\phi<\phi_c$ at $\text{St} > 1$ (and away from $\hat{\dot{\gamma}} = 1$). 
As the Stokes number is decreased, and the fluid increasingly governs the net particle level forces, the lubrication fabric tends to be more isotropic than the contact fabric and the correlation length is increased as particles interact through increasingly strong networks of lubrication films. These conditions move the rheology away from inertial, collisional flow to a viscous flow regime, where the bulk stress is dominated by the fluid contribution.
Above $\phi_c$ there is a jump in the correlation length, combined with a move to a more isotropic microstructure, suggesting the presence of sustained force networks as opposed to collisional rheology. These networks dominate the behavior in the quasistatic regime, resulting in shear rate independent rheology. As $\hat{\dot{\gamma}} \to 1$, the fluid forces, which align with the extensional axis, become significant again, coupled with the move back to rate dependent rheology in the intermediate and soft viscous regimes.


\section{Constitutive model \label{sec:const}}

Following recent discussions that the inertial and viscous numbers can characterize additive contributions to the total suspension stress from contact and fluid effects respectively~\cite{Trulsson2012}, we take inspiration from a recent constitutive model~\cite{Chialvo2012} for steady, simple shear in dry granular media and propose that a similar model for the suspension pressure can be obtained simply by adding a fluid stress contribution (which is a function of the material parameter $\hat{\eta}_f$) to the particle contact stress of the flowing regimes. Chialvo et al.~\cite{Chialvo2012} define $\hat{\text{P}} = \text{P}d/k_n$ separately in each of the dry granular flow regimes; inertial, quasistatic and intermediate. Here we define analogous constitutive regimes, extending upon the previous model to include the $\sigma^F_{xy}$ dominated rheology.  For moderate shear rates below $\phi_c$, we define a hard-particle regime, with characteristic pressure $\hat{\text{P}}_\text{hard}$, that captures the viscous-to-inertial rheology. Above $\phi_c$ we employ the previous model for $\hat{\text{P}}_{\text{QS}}$ directly, as it has been ascertained that the contact stress always dominates in the quasistatic region. For very high shear rates, where $\hat{\dot{\gamma}} \to 1$, we define a soft-particle regime $\hat{\text{P}}_\text{soft}$, capturing the intermediate, shear-thinning behaviour of the previous model as well as the soft viscous behaviour predicted by the above simulations.
The pressure in these regimes is predicted according to the following equations
\begin{subequations}
\small{
\begin{equation}
\hat{\text{P}}_\text{hard} = 
\underbrace{\alpha_\text{hard}^{\text{c}}|\phi - \phi_c|^{\beta_\text{hard}^{\text{c}}} \hat{\dot{\gamma}}^{2}}_\text{contact} 
+
\underbrace{\alpha_\text{hard}^{\text{f}} \hat{\eta}_f|\phi - \phi_c|^{\beta_\text{hard}^{\text{f}}} \hat{\dot{\gamma}}}_\text{fluid} \text{,}
\end{equation}

\begin{equation}
\hat{\text{P}}_\text{QS} = 
\underbrace{\alpha_\text{QS}|\phi - \phi_c|^{\beta_\text{QS}} \hat{\dot{\gamma}}^{0}}_\text{contact}\text{,}
\end{equation}

\begin{equation}
\hat{\text{P}}_\text{soft} = 
\underbrace{\alpha_\text{soft}^{\text{c}}|\phi - \phi_c|^{\beta_\text{soft}^{\text{c}}} \hat{\dot{\gamma}}^{0.5}}_\text{contact}
+
\underbrace{\alpha_\text{soft}^{\text{f}} \hat{\eta}_f|\phi - \phi_c|^{\beta_\text{soft}^{\text{f}}} \hat{\dot{\gamma}}}_\text{fluid}\text{.}
\end{equation}
}
\end{subequations}
It should be noted that the $k_n$ scaling of the constitutive model is maintained in order to capture all of the flow regimes within the same framework. For the hard particle branch, however, the model can be recast to capture the viscous-to-inertial transition as a function of the Stokes number and volume fraction, independently of $k_n$

\begin{equation}
\small{
\frac{{\text{P}}_\text{hard}}{\eta_f \dot{\gamma}} = 
\underbrace{\alpha_\text{hard}^{\text{c}}|\phi - \phi_c|^{\beta_\text{hard}^{\text{c}}} \text{St}}_\text{contact} 
+
\underbrace{\alpha_\text{hard}^{\text{f}} |\phi - \phi_c|^{\beta_\text{hard}^{\text{f}}}}_\text{fluid} }\text{.}
\end{equation}

We demonstrate that the values (of $\alpha_\text{hard}^\text{c}$, $\alpha_\text{QS}$, $\alpha_\text{soft}^\text{c}$, $\beta_\text{hard}^\text{c}$, $\beta_\text{QS}$, $\beta_\text{soft}^{\text{c}}$) proposed by Chialvo \cite{Chialvo2012} in each of the flow regimes are applicable for our contact stress data, and we use the fluid stress data from our simulation results to determine suitable values for the equivalent fluid stress parameters in each regime. 
The value of the shear rate exponent for the contact contribution for each flow regime is consistent with our previous discussion of bulk rheology for the contact dominated regimes: $\hat{\text{P}}_\text{hard} \propto \hat{\dot{\gamma}}^2$; $\hat{\text{P}}_\text{QS} \propto \hat{\dot{\gamma}}^0$; $\hat{\text{P}}_\text{soft} \propto \hat{\dot{\gamma}}^{0.5}$. The fluid contribution to suspension pressure is linear in $\hat{\dot{\gamma}}$ in the hard and soft regimes.

\begin{figure}
  \centering
      \subfigure[]{
  \includegraphics[height=40mm]{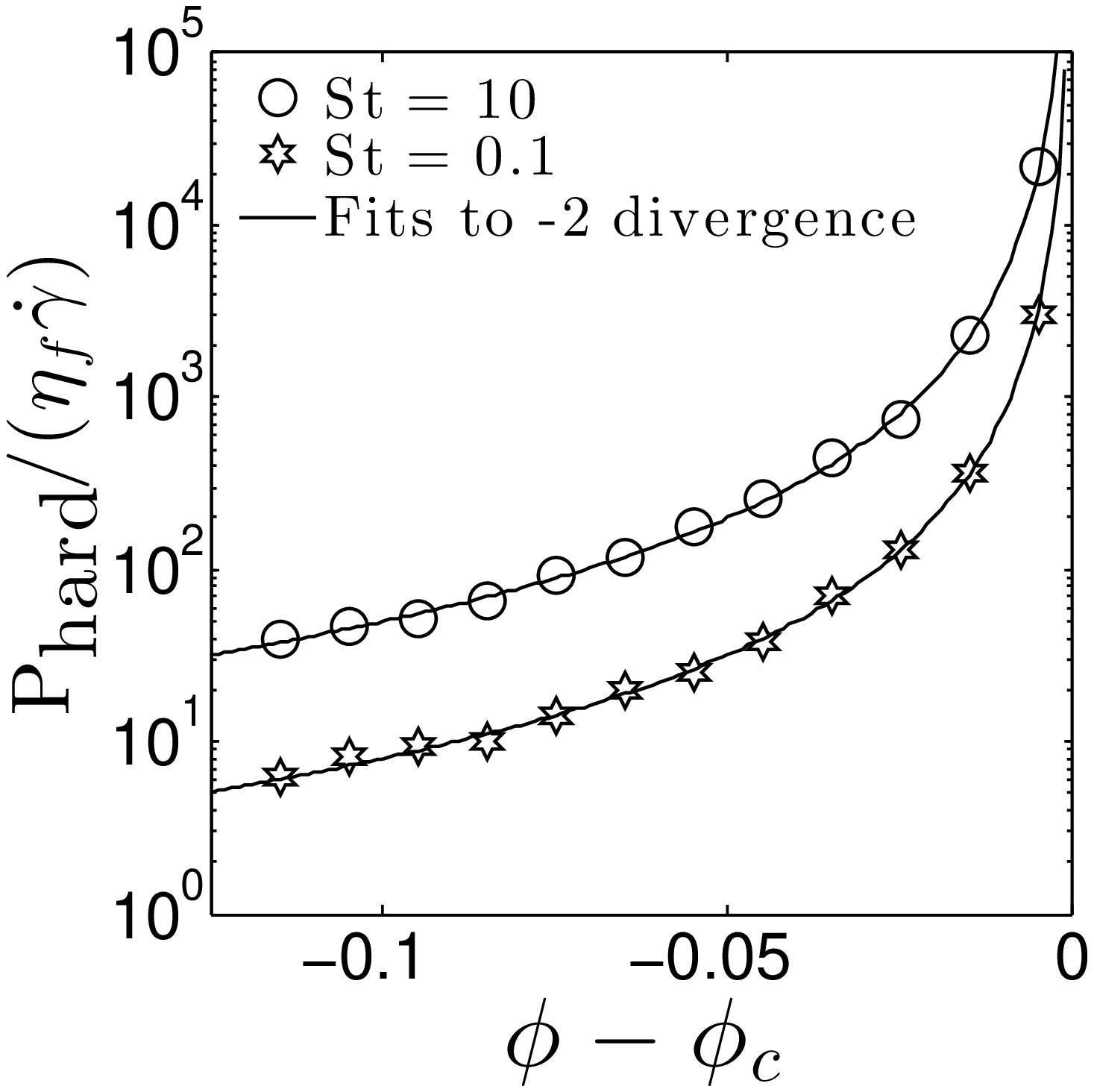}}
      \subfigure[]{
  \includegraphics[height=40mm]{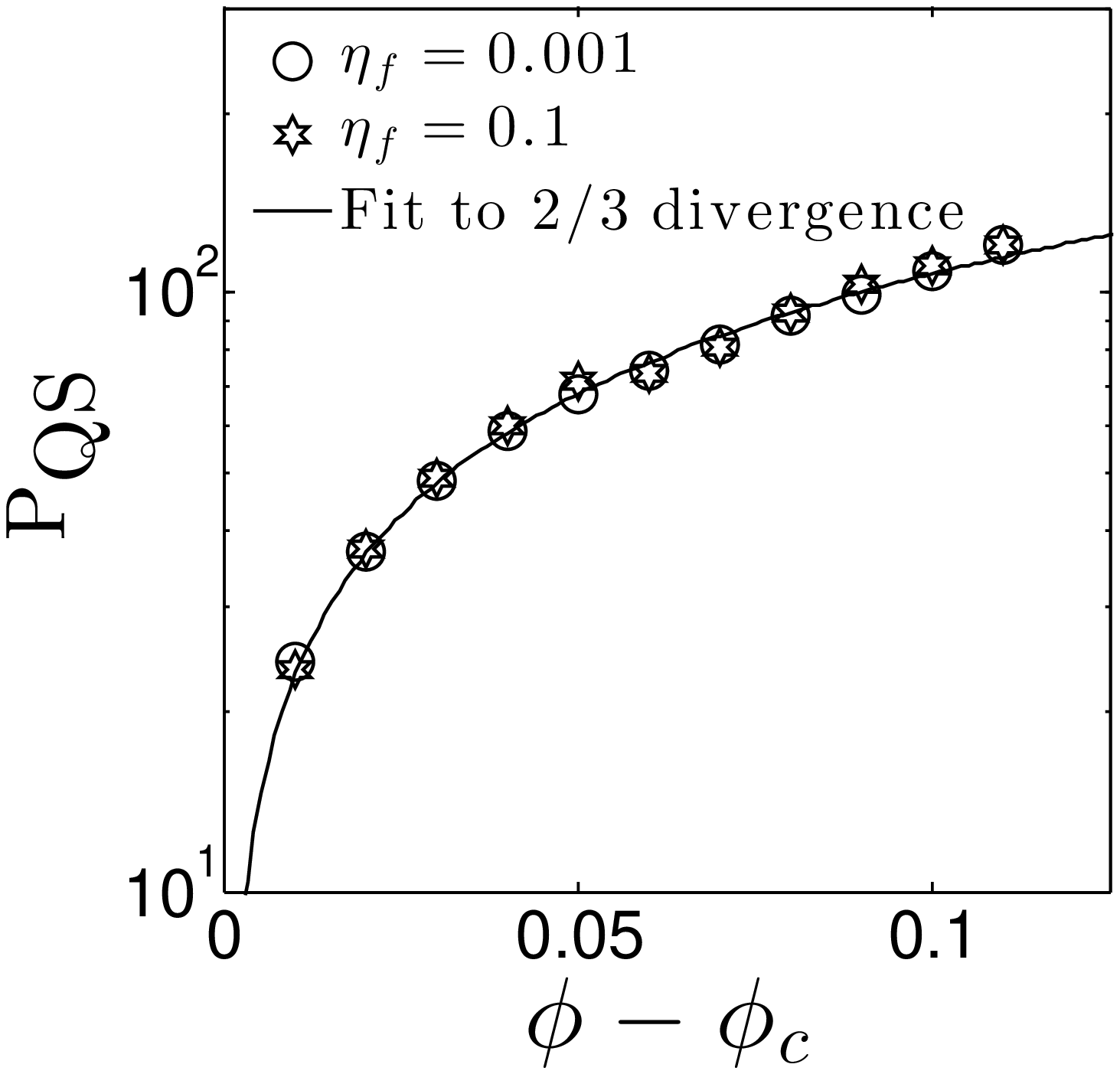}}
     \caption{
     Divergence of pressure around $\phi_c$ in the $\hat{\dot{\gamma}} \ll 1$ limit, demonstrating the forms of the divergence for (a) $\phi<\phi_c$ and (b) $\phi> \phi_c$. Symbols represent simulation results, solid lines represent fits to the constitutive model exponents.
}
 \label{fig:p_divergence}
\end{figure}

The $\beta$ parameter gives the divergence of the contact and fluid pressure contributions with volume fraction. 
As is shown in Section \ref{sec:bulk}, there is no volume fraction dependence of the bulk stress in the soft particle limit as $\hat{\dot{\gamma}} \to 1$. We therefore take $\beta_\text{soft}^\text{c} = \beta_\text{soft}^\text{f} = 0$.
In the hard particle regime, the pressure diverges with $|\phi - \phi_c|^{-2}$ in both the inertial ($\text{St} = 10$) and viscous ($\text{St}=0.1$) cases, as shown in Figure~\ref{fig:p_divergence}a. This leads us to set $\beta_\text{hard}^c = \beta_\text{hard}^f = -2$. The pressure is notably higher for the inertial case, consistent with the shear thickening associated with Bagnoldian stress scaling in this flow regime.
 In the quasistatic regime, the pressure increases above $\phi_c$ according to $|\phi - \phi_c|^{\frac{2}{3}\pm 0.01}$, independently of the fluid viscosity (Figure~\ref{fig:p_divergence}b). It is not surprising that this behavior is not dependent on the fluid properties, since we have already concluded that at very low shear rate and high volume fraction, i.e. in the quasistatic regime, the flow is always contact dominated for the range of fluid viscosities studied here. This further justifies the lack of a fluid contribution to $\hat{\text{P}}_{\text{QS}}$ in our model. A value of $\beta_\text{QS} =2/3$ is used for the constitutive model.

The multiplicative fitting parameters $\alpha$ are obtained from Chialvo et al. \cite{Chialvo2012}~for the contact terms, noting that $\alpha_\text{QS}$ is dependent on the choice of friction coefficient in the contact model. In addition, we obtain $\alpha_\text{hard}^\mathrm{f} = 0.02$ and $\alpha_\text{soft}^\mathrm{f} = 0.55$. A summary of all model parameters is given in Table 1.

A blending function is employed, identical to that proposed by Chialvo et al.~\cite{Chialvo2012}, to combine the individual contributions from the limits of each flow regime, giving the total pressure $\hat{\text{P}}$ as a function of $\hat{\dot{\gamma}}$, $\hat{\eta}_f$ and $|\phi-\phi_c|$ across all flow regimes:
\begin{equation}
\hat{\text{P}} = \left\{ 
  \begin{array}{l l}
    \hat{\text{P}}_{\text{QS}} + \hat{\text{P}}_{\text{soft}} & \quad \phi \geq \phi_c\\
    \\
    (\hat{\text{P}}_{\text{hard}}^{-1} + \hat{\text{P}}_{\text{soft}}^{-1})^{-1} & \quad \phi < \phi_c 
  \end{array} \right. \text{.}
  \end{equation} 
The critical volume fraction for granular jamming $\phi_c$, known to be a function of the particle-particle friction coefficient ($\mu_p$) and the extent of bidispersity, has been determined for the present case at both macro- and microscopic levels. At the macroscopic level, a transition from viscous (or inertial) to quasistatic flow is observed for $\hat{\dot{\gamma}} \ll 1$ between $\phi = 0.57$ and $\phi=0.59$. Furthermore, by setting $\phi_c = 0.585$ the divergence of suspension pressure with $\phi-\phi_c$ is effectively captured as $|\phi-\phi_c| \to 0$. At the micro-scale, the velocity correlation length is observed to diverge between $\phi = 0.58$ and $\phi = 0.59$. The value of $\phi_c$ used in this case is therefore 0.585, slightly higher than that ($\phi_c = 0.581$) reported for monodisperse particles of the same friction coefficient, as expected.
  
  \begin{table}
\begin{tabular}{ c c c c c c c c c c }
$\alpha_\text{hard}^c$ &
$\alpha_\text{QS}$ &
$\alpha_\text{soft}^c$ &
$\alpha_\text{hard}^f$ &
$\alpha_\text{soft}^f$ &
$\beta_\text{hard}^c$ &
$\beta_\text{QS}$ &
$\beta_\text{soft}^c$ &
$\beta_\text{hard}^f$ &
$\beta_\text{soft}^f$\\ \hline
0.021 &
0.25 &
0.099 &
0.02 &
0.55 &
-2 &
2/3 &
0 &
-2 &
0 \\ \hline
\end{tabular}
\caption{The parameters used in the constitutive model.}
\end{table}

\begin{figure}
  \centering
  \includegraphics[height=60mm]{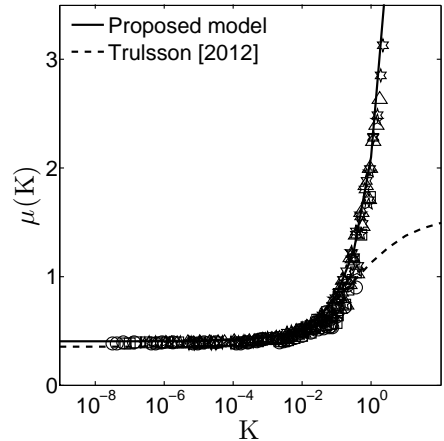}
     \caption{
Stress ratio $\mu$ as a function of $\text{I}_\text{I}$ and $\text{I}_\text{V}$. Different symbols represent different shear rates ($10^{-5}$ to $10^{-2}$), fluid viscosities ($10^{-5}$ to $10^{-2}$) and volume fractions (0.48 to 0.68). The solid line represents the constitutive model in Equation \ref{stress_ratio_model}; the dashed line represents a constitutive model proposed previously by Trulsson et al.~\cite{Trulsson2012}.
 }
  \label{fig:stress_ratio}
\end{figure}

In order to calculate the shear stress from the pressure, we adopt the popular $\mu(\text{I}_\text{I})$ rheology and appeal to a recent constitutive model for the stress ratio $\mu = {\sigma}_{xy}/{\text{P}}$ as a function of $\text{K} = \text{I}_\text{V} + \alpha \text{I}_\text{I}^2$~\cite{Trulsson2012}, combining inertial and viscous rheology. We find that the proposed value of $\alpha = 0.635 \pm 0.009$ is suitable for $\text{K}<10^{-2}$, but that a value of $\alpha=0.3$ allows the rheological contributions to be combined successfully for $\text{K} < 1$. Furthermore, the proposed function $\mu({\text{K}})$ does not take the correct form for large $\text{I}_\text{V}$. As demonstrated by the experimental results and constitutive model given by Boyer \cite{Boyer2011a}, $\mu$ is expected to diverge as $\text{I}_\text{V} \to \infty$ at $\text{I}_\text{I} = 0$, while the model of Trulsson et al \cite{Trulsson2012} predicts a maximum in $\mu$. We therefore propose a modified form that captures the divergence in $\mu$ with $\text{K} \to \infty$,
\begin{equation}
\mu(\text{K}) = 0.38 + 1.2 \text{K} ^ {1/2} + 0.5 \text{K} \text{.}
\label{stress_ratio_model}
\end{equation}
It is found that all the simulation data can be described by this model, with the exception of extremely high shear rates ($\hat{\dot{\gamma}} > 0.1$) where particle overlaps are expected to become unfeasibly large compared to that of granular mateirals under normal experimental conditions. Stress ratio data are given in Figure~\ref{fig:stress_ratio}, along with predictions given by the present model, and that proposed previously.
We note that previous $\mu(\text{I})$ models \cite{DaCruz2005,Boyer2011a} for dry granular materials predict a maximal $\mu$ for $\text{I}_\text{I}\to \infty$ at $\text{I}_\text{V} = 0$, while the present model diverges for $\text{I}_\text{I} \to \infty$. Experimentally, the inertial number is not observed to exceed a value of around 0.3, so we conclude that our model does capture the $\mu$ dependence on $\text{I}_\text{I}$ at $\text{I}_\text{V}=0$ within the experimentally accessible range, as well as the experimentally observed divergent behavior as $\text{I}_\text{V} \to \infty$.

Constitutive model predictions are given as the solid line in Figure~\ref{fig:flowcurve}, demonstrating good agreement with the simulation results. Furthermore, we note the ability of the model to capture the divergence of the suspension viscosity with volume fraction in the viscous, inertial and quasistatic regimes. Below $\phi_c$ the suspension viscosity diverges as $|\phi - \phi_c|^{-2}$ in the viscous and inertial regimes, consistent with the divergence of the bulk pressure plotted above. This result is consistent with the experimental and simulation results in~\cite{Kawasaki2014,Mari2014a,Boyer2011a} and also with the traditionally cited Quemada equation~\cite{Quemada1978}, $\mu = (1 - \phi/\phi_c)^{-2}$. 
It has also been found that by excluding the tangential lubrication force from the present simulations, the exponent tend towards -1 for the viscous regime, which is consistent with the early theoretical derivation by Frankel and Acrivos~\cite{Frankel1967} assuming a purely normal lubrication interaction. 
There is still debate as to the true nature and form of this divergence, with the consensus being that the exponent is somewhere between -1 and -3~\cite{Kawasaki2014a}.
Above $\phi_c$, the viscosity and scales with $|\phi - \phi_c|^{1}$, giving some agreement with experimental work~\cite{Nordstrom2010a} that finds a value of around 1.35 in this regime.

\section{Conclusions and discussions \label{sec:conc}}
The particle dynamics of dense granular suspensions have been simulated using a discrete element method combining particle contact and hydrodynamic lubrication. Simulations of homogeneous simple shear flow have been performed, shedding light on the transitions between flow regimes as a function of solid volume fraction ($\phi$), shear rate ($\dot{\gamma}$) and material properties ($\hat{\eta}_f$). 
We found that for volume fractions below a critical value, quasi-Newtonian behaviour emerges at Stokes numbers below 1, transiting to the continuously shear thickening, Bagnoldian behavior above 1. A quasistatic, rate-independent regime exists above $\phi_c$ for low and moderate shear rates. At very high shear rates, which we have defined as $\hat{\dot{\gamma}} \to 1$, the flow becomes $\phi$-independent and either shear thinning or viscous, depending on the value of a material parameter, $\hat{\eta}_f$.
All the transitions are shown to correlate with a change in the relative importance of the lubrication contacts at the microscopic scale and their contribution to the total stress at the macroscopic scale.

The transitions in bulk rheology are well correlated with changes in microstructure, characterized by distributions of interacting forces and relative velocities, fabric and correlation length. When the viscous effect is strong, the force distribution and fabric are more isotropic and the correlation length is longer. Interestingly, the force distribution and fabric characterizing mechanical contacts behave distinctly from those for the lubrication contacts. They remain anisotropic with the major principal direction aligned with the compressive axis while the lubrication contacts become more isotropic or flip the major principal direction to the extensional axis. Although the direct consequence on the bulk rheology is not clear, this distinction between the two different contact networks might have important implications for modeling more complex unsteady rheology of dense granular suspensions.

With such understanding of the rheological behavior, constitutive equations for pressure have been established for the asymptotic flow behaviors using an additive form combining viscous effects with dry granular rheology. The equations are then bridged ad hoc using a blending function to capture the transitions between them. The shear stress-to-pressure ratio is modeled as a function of both the inertial and the viscous numbers for all flow regimes, with a form applicable to a wider parametric range than previous models. The resultant constitutive model has been shown to be able to capture all the flow curves from the DEM simulations and can predict the divergent behavior of the suspension viscosity with respect to the solid volume fraction. The current model, with only scalar representation of stress and strain rate calibrated with data from simple shear flow, is not expected to capture different types of flow, e.g.~extensional flow. Future work is warranted to generalize to a fully tensorial model supported with data from simulation and experiments of different types of flows. 
\section*{Acknowledgements}

This work is funded by EPSRC and Johnson Matthey through a CASE studentship award. The authors would like to thank Joe D. Goddard for pointing to the paper by Frankel and Acrivos, and Jin Y. Ooi, Wilson C.K. Poon, Ben Guy, Michiel Hermes, Michael E. Cates, Paul McGuire, Michele Marigo, Hugh Stitt and Han Xu for helpful discussions.

\bibliography{CJN_paper.bib}
    \bibliographystyle{ieeetr}

\end{document}